\documentclass{article}

\usepackage{arxiv}

\usepackage[utf8]{inputenc} 
\usepackage[T1]{fontenc}    
\usepackage{lmodern}        
\usepackage{hyperref}       
\usepackage{url}            
\usepackage{booktabs}       
\usepackage{amsfonts}       
\usepackage{nicefrac}       
\usepackage{microtype}      
\usepackage{graphicx}

\hypersetup{colorlinks,
            linkcolor=[rgb]{0.067,0.467,0.200}, 
            citecolor=[rgb]{0.533,0.180,0.447}, 
            urlcolor=[rgb]{0.000,0.267,0.533}} 

\title{Bayesian calibration of simulation models}
\subtitle{A tutorial and an Australian smoking behaviour model}

\author{
\small%
 \mbox{\href{https://orcid.org/0000-0002-2573-9683}{\includegraphics[scale=0.06]{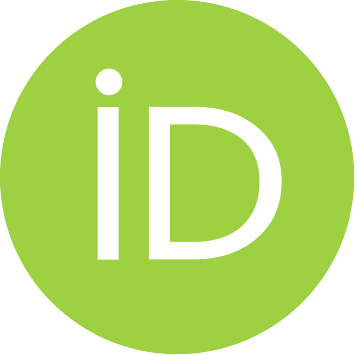}}} %
Stephen Wade\\
\small  The Daffodil Centre, 
The University of Sydney\thanks{A Joint Venture with Cancer Council NSW}\\
 \small Kings Cross, NSW 1340 \\ %
 \small \texttt{\href{mailto:stephen.wade@sydney.edu.au}{\nolinkurl{stephen.wade@sydney.edu.au}}} %
 \And
\small%
 \mbox{\href{https://orcid.org/0000-0001-5731-9651}{\includegraphics[scale=0.06]{orcid.pdf}}} %
Marianne F Weber\\
\small  The Daffodil Centre, 
The University of Sydney\footnotemark[\value{footnote}]\\
 \small Kings Cross, NSW 1340 \\ %
 \And
\small%
 \mbox{\href{https://orcid.org/0000-0001-9596-6825}{\includegraphics[scale=0.06]{orcid.pdf}}} %
Peter Sarich\\
\small  The Daffodil Centre, 
The University of Sydney\footnotemark[\value{footnote}]\\
 \small Kings Cross, NSW 1340 \\ %
 \And
\small%
 \mbox{\href{https://orcid.org/0000-0001-9213-8733}{\includegraphics[scale=0.06]{orcid.pdf}}} %
Pavla Vaneckova\\
\small  The Daffodil Centre, 
The University of Sydney\footnotemark[\value{footnote}]\\
 \small Kings Cross, NSW 1340 \\ %
 \And
\small%
 \mbox{\href{https://orcid.org/0000-0003-2287-8220}{\includegraphics[scale=0.06]{orcid.pdf}}} %
Silvia Behar-Harpaz\\
\small  The Daffodil Centre, 
The University of Sydney\footnotemark[\value{footnote}]\\
 \small Kings Cross, NSW 1340 \\ %
 \And
\small%
 \mbox{\href{https://orcid.org/0000-0001-5453-0734}{\includegraphics[scale=0.06]{orcid.pdf}}} %
Preston J Ngo\\
\small  The Daffodil Centre, 
The University of Sydney\footnotemark[\value{footnote}]\\
 \small Kings Cross, NSW 1340 \\ %
 \And
\small%
 \mbox{\href{https://orcid.org/0000-0003-4769-8082}{\includegraphics[scale=0.06]{orcid.pdf}}} %
Sonya Cressman\\
\small  Faculty of Health Sciences, 
Simon Fraser University\\
 \small Burnaby, BC V5A 1S6 \\ %
 \And
\small%
 \mbox{\href{https://orcid.org/0000-0002-6651-8035}{\includegraphics[scale=0.06]{orcid.pdf}}} %
Coral E Gartner\\
\small  School of Public Health, 
The University of Queensland\\
 \small Brisbane St Lucia, QLD 4072 \\ %
 \And
\small%
 \mbox{\href{https://orcid.org/0000-0001-9314-2283}{\includegraphics[scale=0.06]{orcid.pdf}}} %
John M Murray\\
\small  School of Mathematics and Statistics, 
UNSW\\
 \small Sydney, NSW 2052 \\ %
 \And
\small%
 \mbox{\href{https://orcid.org/0000-0002-6995-4369}{\includegraphics[scale=0.06]{orcid.pdf}}} %
Tony A Blakely\\
\small  Melbourne School of Population and Global Health, 
The University of Melbourne\\
 \small Parkville, Victoria 3010 \\ %
 \And
\small%
 \mbox{\href{https://orcid.org/0000-0002-4617-1302}{\includegraphics[scale=0.06]{orcid.pdf}}} %
Emily Banks\\
\small  National Centre for Epidemiology and Population Health, 
The Australian National University\\
 \small Canberra, ACT 2601 \\ %
 \And
\small%
 \mbox{\href{https://orcid.org/0000-0002-4989-5058}{\includegraphics[scale=0.06]{orcid.pdf}}} %
Martin C Tammemagi\\
\small  Faculty of Applied Health Sciences, 
Brock University\\
 \small St.~Catharines, ON L2S 3A1 \\ %
 \And
\small%
 \mbox{\href{https://orcid.org/0000-0002-6443-6618}{\includegraphics[scale=0.06]{orcid.pdf}}} %
Karen Canfell\\
\small  The Daffodil Centre, 
The University of Sydney\footnotemark[\value{footnote}]\\
 \small Kings Cross, NSW 1340 \\ %
 \And
\small%
 \mbox{\href{https://orcid.org/0000-0002-5439-6552}{\includegraphics[scale=0.06]{orcid.pdf}}} %
Michael Caruana\\
\small  The Daffodil Centre, 
The University of Sydney\footnotemark[\value{footnote}]\\
 \small Kings Cross, NSW 1340 \\ %
}

\providecommand{\tightlist}{%
  \setlength{\itemsep}{0pt}\setlength{\parskip}{0pt}}

\usepackage{booktabs,array}
\usepackage{calc}
\usepackage{amsmath}
\usepackage{pdflscape}
\newcommand{\blandscape}{\begin{landscape}}
\newcommand{\elandscape}{\end{landscape}}
\usepackage{rotating}
\usepackage[numbers,square]{natbib}
\usepackage[labelfont=bf]{caption}

\usepackage{afterpage}
\usepackage{flafter}
\fancypagestyle{lscape}{%
 
}

\begin{document}
\maketitle

\newpage
\begin{abstract}
Simulation models of epidemiological, biological, ecological, and environmental processes are increasingly being calibrated using Bayesian statistics. The Bayesian approach provides simple rules to synthesise multiple data sources and to calculate uncertainty in model output due to uncertainty in the calibration data. As the number of tutorials and studies published grow, the solutions to common difficulties in Bayesian calibration across these fields have become more apparent, and a step-by-step process for successful calibration across all these fields is emerging. We provide a statement of the key steps in a Bayesian calibration, and we outline analyses and approaches to each step that have emerged from one or more of these applied sciences. Thus we present a synthesis of Bayesian calibration methodologies that cut across a number of scientific disciplines.
To demonstrate these steps and to provide further detail on the computations involved in Bayesian calibration, we calibrated a compartmental model of tobacco smoking behaviour in Australia. We found that the proportion of a birth cohort estimated to take up smoking before they reach age 20 years in 2016 was at its lowest value since the early 20\textsuperscript{th} century, and that quit rates were at their highest. As a novel outcome, we quantified the rate that ex-smokers switched to reporting as a `never smoker' when surveyed later in life; a phenomenon that, to our knowledge, has never been quantified using cross-sectional survey data.
\end{abstract}

\hypertarget{introduction}{%
\section{Introduction}\label{introduction}}

There are many approaches to the calibration of epidemiological, biological,
environmental, or ecological simulation models. Where data are abundant and
model-parameters are identifiable with such data, or where uncertainty is not of
interest, almost any calibration framework will provide an acceptable point
estimate. What, then, to make of situations where there are less data, and
significant uncertainty exists that is important either as a research question
itself or in decision-making? Consider a government health department using a
model of a screening program to inform a decision on screening frequency; if the
estimate of harm at a given screening frequency is highly uncertain, a
policy-maker may prefer a more conservative frequency depending on the
(modelled) marginal effect on disease-burden. Statement and minimisation of
uncertainty is vital in this and similar contexts, and a Bayesian approach to
calibration provides simple rules to quantify uncertainty \citep{OHagan2006} and
reduce uncertainty by incorporating all available data \citep{Jackson2015, Menzies2017, Birrell2018}.

We are not the first to write a tutorial on Bayesian model calibration and its
benefits over conventional fitting procedure-like methods. Past tutorials reveal
its widespread use, and have collectively described both generic-model
uncertainty analysis \citep{OHagan2006}, and model calibration in health policy
\citep{Jackson2015, Menzies2017} and systems biology \citep{Collis2017}. These fields
benefit from the ability to synthesise disparate data sources, using graphical
models to both map out the dependence structure of the data \citep{Jackson2015, Wilkinson2007} and to identify a procedure to sample (posterior) values of
parameters via languages such as BUGS \citep{Lunn2012}. Analysis of identifiability
of parameters has emerged as a theme in successful calibration \citep{Alahmadi2020, Alarid-Escudero2018}, and Bayesian-specific approaches to such analyses have
emerged \citep{Raue2013}. Furthermore, Bayesian calibration may succeed where other
approaches fail through greater use of prior information.

The aim of this tutorial was to strengthen connections between existing
knowledge on Bayesian calibration from systems biology, epidemiology, ecology,
infectious disease modelling, and health economic evaluation. In this tutorial
we outline the common steps and tools used across these fields in successful
Bayesian calibration of models and we discuss how to use the results of a
calibration to make statements of uncertainty.

We demonstrated these steps by calibrating a model of tobacco smoking
behaviour. The model was a generalisation of the model of Gartner et al.
\citep{Gartner2009} to a continuous-in-time non-Markov compartmental model, governed
by delay differential equations. This example highlighted how a model of a risk
factor from epidemiology, governed by a class of equations more familiar to
systems biologists (where they have been used to model breast cancer cell
growth \citep{Simms2012}) can be calibrated using Bayesian statistics. Using the
calibrated model, we described the value and uncertainty of the model's;
proportion that initiated smoking in birth cohorts born between 1910 and 1996;
rate of quitting smoking between 1930 and 2016, and; rate that ex-smokers
switched to reporting as a never smoker when surveyed later in life. This final
quantity has never been quantified using cross-sectional smoking survey data.

\hypertarget{bayesian-calibration-of-simulation-models-tutorial}{%
\section{Bayesian calibration of simulation models: Tutorial}\label{bayesian-calibration-of-simulation-models-tutorial}}

A mathematical simulation model, which we refer to simply as a `model', is an
abstract representation of a process (or system) in which a parameterised set of
equations transform inputs into outputs or predictions. Model calibration is a
procedure by which values of the parameters are selected. Calibration involves
taking measurements of inputs and outputs from studies of the process and then
computationally searching for parameter values for which the model predictions
closely match the observations.

In a Bayesian calibration, both the parameters and model predictions are defined
as random variables. Calibration is reframed as the calculation of the posterior
distributions of these random variables given the observed data. The posterior
distribution, or simply `the posterior', describes the uncertainty owing to the
randomness of the collected data. The knowledge of the parameters prior to data
collection is described by the prior distribution, or simply, `the prior'.

We use the following notation:

\begin{itemize}
\tightlist
\item
  Capital letters, such as \(\Theta\) and \(X\), represent real-valued random
  variables, specific values are denoted with lower case letters such as
  \(\theta\) and \(x\).
\item
  The probability density (or mass) function of a random variable is denoted
  as \(f_\Theta(\theta)\).
\item
  The random variable \(\Theta\) conditional on \(X\) is denoted \(\Theta | X\),
  with density (or mass) denoted as \(f_{\Theta | X}(\theta | x)\).
\item
  A function's argument may be suppressed when it is obvious, e.g., \(f_\Theta\)
  instead of \(f_\Theta(\theta)\).
\end{itemize}

Denoting the parameters of the model as \(\Theta\) and the observed data as \(X\),
then; the prior is given by \(f_\Theta\); the posterior is the random variable
\(\Theta | X\); and these are related to each other through the likelihood
function \(\mathcal{L}_{X | \Theta}\) by Bayes' rule:

\[ f_{\Theta | X} = \frac{ \mathcal{L}_{X | \Theta}  f_\Theta }{ f_X }.  \]

Although unfamiliar to non-Bayesian or non-statistics trained modellers,
there are similarities between `fitting procedure' approaches to calibration and
the Bayesian approach. Similarly to Vanni et al. \citep{Vanni2011} we describe
Bayesian calibration in seven steps:

\begin{enumerate}
\def\labelenumi{\arabic{enumi}.}
\tightlist
\item
  Select the parameters to calibrate.
\item
  Select a joint prior of the parameters given available prior knowledge.
\item
  Derive a statistical model of the `observation processes' used to collect
  data.
\item
  Assess a region of the parameter space to sample from, with possible return
  to steps (1) or (2).
\item
  Sample from the posterior distribution of the parameters.
\item
  Refine or select the model, using model fit statistics, with possible return
  to step (1).
\item
  Use the posterior sample to describe predictions and uncertainty.
\end{enumerate}

These steps are not prescriptive; however, this structure may reduce a
modeller's workload by providing both early warning of failure and potential
remedies. Next, we will discuss each of these steps in more detail.

\hypertarget{select-the-parameters-to-calibrate}{%
\subsection{Select the parameters to calibrate}\label{select-the-parameters-to-calibrate}}

Our first step in any calibration, Bayesian or otherwise \citep{Vanni2011}, is to
determine for which parameters will values be estimated. For this we should
investigate \emph{structural identifiability}, which is a property of the model,
some ideal input and output, and the parameters. If a set of parameters is
structurally identifiable, then one and only one distribution of the
parameter-values is allowed for any input and output of the given ideal form.
We loosely paraphrase this by saying structural identifiability means the set of
parameters can be uniquely determined by the ideal input and output (more
formal discussion of identifiability and the related concept of parameter
redundancy can be found in Cole 2020 \citep{Cole2020}). Structural identifiability is
a necessary (but insufficient) condition for a more important characteristic
called \emph{practical identifiability}, i.e.~that the set of parameters can be
uniquely determined by the observed calibration data.

Structurally non-identifiable parameters cannot be calibrated with data only
from observed output of the modelled process; other data sources are necessary.
For example, if the cell division rate is a parameter of the model, but it is
not identifiable, then some other experimental study, direct measurement,
published estimate, or expert-elicited estimate is required to calibrate the
model.

Formal methods to determine structural identifiability have been developed in
ecology \citep{Cole2020} and systems biology \citep{Chis2011, Miao2011}, although uptake
is limited in the latter field \citep{Nguyen2016} and in fields that use similar
mathematical models such as infectious disease modelling and epidemiology
\citep{Alahmadi2020}. Nonetheless, any investigation of structural identifiability
can guide us towards which additional data will be required other than
observations of the process. \emph{A priori} practical identifiability is difficult
to establish \citep{Alahmadi2020} and, as such, practical identifiability is assessed
in a later step with observed calibration data and sometimes using simulated
data.

\hypertarget{select-a-prior}{%
\subsection{Select a prior}\label{select-a-prior}}

We assign a prior, \(f_\Theta\), to the parameters which reflects the available
prior knowledge about their value. Priors can be either `informative', in which
case they quantify established certainty that the parameter takes one value over
another, or `non-informative', where the intention is that the corresponding
posterior reflects only the data used in calibration. Informative priors can be
estimated from literature, taken directly from a previous analysis, or the
result of an elicitation procedure, and must be obtained independently of the
calibration data. Non-informative priors are designed to fulfil a chosen set of
conditions or optimise some abstract criteria, for example the Jeffreys prior
always satisfies a criterion of invariance under any change of variables.

Structural identifiability is an important consideration when it comes to the
choice of informative versus non-informative prior. If structural
identifiability has been demonstrated we are free to choose from either
informative or non-informative priors depending on; our preference, the context,
and feasibility \citep{Goldstein2006}. If the model is to be used to evaluate a novel
pharmacotherapy, selecting non-informative priors may reduce the perception of
bias or conflict of interest compared to selecting informative (subjective)
priors \citep{Berger2006}. On the other hand, informative priors can reduce
uncertainty and improve decision making. For parameters that are structurally
non-identifiable, an informative prior may be the only option, otherwise
algorithms used in analysis of practical identifiability and for sampling from
the posterior in later steps are unlikely to converge.

A non-informative prior might be \emph{improper}, which means that the integral
\(\int_\Theta f_\Theta(\theta) \, \mathrm{d} \theta\) is not finite, and it is
not a true `distribution'. It is irrelevant to calibration whether the prior
is proper, however, the posterior must be. We should demonstrate that the
resulting posterior is proper, however this may be difficult to the point of
impracticality. While a proper posterior may guarantee convergence,
seemingly-converged output from the sampling algorithms we discuss below is no
guarantee that the posterior is proper \citep{Hobert1996}. Any evidence of lack
of convergence is, at the least, discouraging, and we discuss this further in
the sampling step.

\hypertarget{derive-a-statistical-model-of-the-observation-processes}{%
\subsection{Derive a statistical model of the observation processes}\label{derive-a-statistical-model-of-the-observation-processes}}

In conventional calibrations the observation processes by which the calibration
data were collected may not elicit much modelling interest. These processes can
include experiments, surveys, a Census, scraping from websites or apps, or some
other type of study. In Bayesian calibration we assume each observation process
is a random process, and the collected data are viewed as one outcome. With few
exceptions, such as `Approximate Bayesian Computation' \citep{Beaumont2010}, we must
construct a statistical model of the observation process. This requirement is
not exclusive to Bayesian calibration; the likelihood of the observed data,
given a statistical model thereof, is one option for the objective function in
fitting procedure-based calibrations.

Other than formulating the statistical model and a means of evaluating its
likelihood, we must also construct a map between the model parameters and the
likelihood's typical parameters \citep{Rutter2009, Menzies2017, Alarid-Escudero2018}. As an example, consider a study where the observed data
were outcomes of independent binomial processes. We would make a map between
the model's parameters, conditioned on covariates that were assumed fixed for
each observation, to the probabilities of success for each observed binomial
trial. In more concrete mathematical notation (suppressing notation for
covariates); given the random variable \(\Phi\), which is a vector of success
probabilities for each observation, we would construct a map \(M : \Theta \to \Phi\). The likelihood of the data \(x\) given model parameters \(\theta\) would be;

\[
    \mathcal{L}_{X | \Phi} (X=x | \Phi=M(\theta)) = %
        \prod_{i} \binom{ n_i }{ x_i } %
            { M_i(\theta) }^{ x_i } (1 - M_i(\theta))^{ n_i - x_i },
\]

where the subscript \(i\) refers to the i\textsuperscript{th} component of a vector, and \(n\) is a
vector of the number of trials from the observation process. The right-hand
side is the product of the likelihood of independent Binomial trials (given
our model \(M\)).

\begin{figure}
\centering
\includegraphics{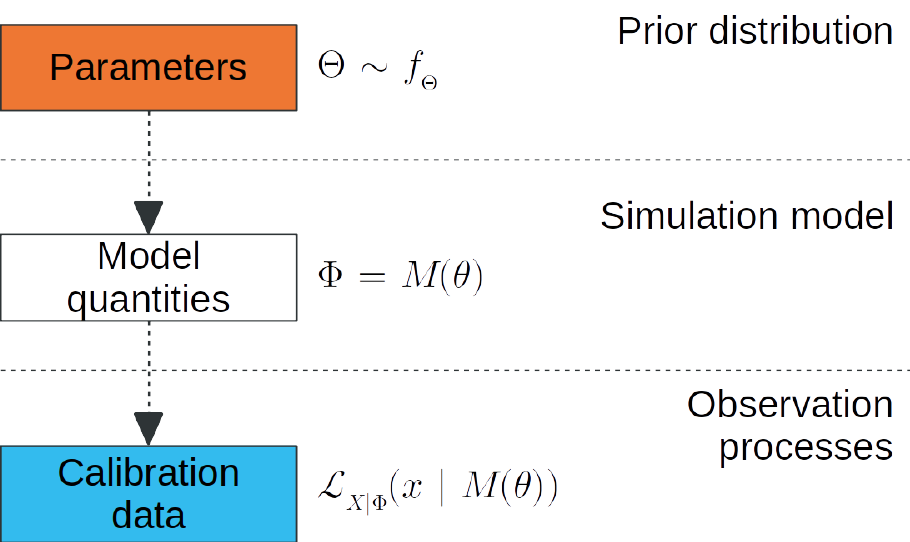}
\caption{\label{fig:graphical-model}Directed acyclic graph outlining the conditional dependency of generic quantities in a Bayesian calibration. The direction of the arrow indicates parent-child relationships. A node is independent of all nodes conditioned on its parent and descendants. Model parameters, denoted \(\Theta\), have marginal distribution given by the prior, \(f_\Theta\), and the model provides a map, \(M\), from the model parameters to those of the likelihood of the observed data i.e.~\(\mathcal{L}_{X | \Phi}( X | \Phi=M(\theta) ) \equiv f_{X | \Theta}\).}
\end{figure}

In this example, each observation was assumed independent of the others, but
this is not always valid, particularly when using multiple sources of data. For
example, cancer incidence recorded in a population registry and population
cancer mortality rates via death certificates are clearly not independent of
each other. In infectious disease modelling, dependence can arise when some
individuals are observed in multiple datasets; Alahmadi et al. \citep{Alahmadi2020}
provided an example in influenza outbreak modelling where the same people could
be captured in a `first few hundred' study as well as in case-notification
systems. When the dependence structure of the observation processes is
non-trivial, we can use a directed acyclic graph \citep{Lunn2012} (DAG) to both
define and visualise the structure of the random variables and the data in
calibration \citep{Jackson2015}. In these graphs, nodes represent quantities in the
calibration, and directed edges represent parent-child relationships, such that
any node is independent of all other nodes conditional on its parent and
descendants \citep{Lunn2012}. The model-parameter nodes have no parents and are
(marginally) distributed as per the prior; the observed-data nodes have no
children, and are distributed as per the observation process from which we have
the likelihood; and quantities `within' the model (which make up the map \(M\))
sit in-between the two former types of node. These graphs can be generated using
the BUGS language \citep{Lunn2012} and at their most generic, have the structure
shown in \textbf{Figure \ref{fig:graphical-model}}.

Biases may also feature in observation processes. For example, a survey may
over-sample a sub-population, or a trial may be subject to a `healthy volunteer'
effect. It may not be strictly better to include additional low-quality data as
opposed to using only high-quality data sources \citep{Ades2006, Moss2017}. We can
include conflict or bias between two data sources in the statistical model and
estimate these as part of the calibration procedure. Jackson et al.
\citep{Jackson2015} calibrated a human papillomavirus prevalence model using data
from both a trial and a population screening registry, which did not agree. They
estimated the discrepancy between the two as part of the calibration and this
had a marked effect on estimates of disease progression rates when compared to
calibrating under an assumption of zero-discrepancy. Lastly, data may be
missing `not-at-random', and a model of missing responses may also be included
and estimated in Bayesian calibration.

\hypertarget{assess-regions-of-the-parameter-space-to-sample-from}{%
\subsection{Assess regions of the parameter space to sample from}\label{assess-regions-of-the-parameter-space-to-sample-from}}

A suitable region of the parameter space to sample from is one where \emph{practical
identifiability} holds, as there is less risk of bias or non-convergence in
the next step. Practical identifiability is the property that the distribution
of the parameter-values can be uniquely determined by the observed calibration
data. It has been further categorised as either `global' or `local', with the
respective meanings that the identified value is unique among all possible
values, or it is unique only within some neighbourhood of itself. We focus on
methods to investigate local identifiability as they can be readily applied.
Practical identifiability should be investigated with either the observed data,
or using simulated data following the examples of Rutter et al. \citep{Rutter2009} or
Alarid-Escudero et al \citep{Alarid-Escudero2018}.

One approach to analyse local identifiability of parameter estimates is to use
the profile likelihood \citep{Raue2009} or the profile posterior \citep{Raue2013}.
Raue et al. \citep[see][\S2.2-2.3]{Raue2009} defined (local) practical
identifiability as; a parameter estimate is identifiable when its likelihood (or
posterior)-based confidence interval (CI) is bounded given a pre-specified
confidence level. Estimates such as the maximum likelihood estimate (MLE) or the
maximum a posteriori (MAP) estimate are most useful \citep{Menzies2017} particularly
if the likelihood or posterior are unimodal. Estimating the profile is
computationally feasible (see \S \protect\hyperlink{numerical-evaluation-of-profile-posterior}{Numerical evaluation of profile
posterior} in Supplementary Material
1) for a small number of parameters but will suffer from the `curse of
dimensionality' in models with many parameters \citep{Alarid-Escudero2018}. For large
numbers of parameters, it may be more practical to consider sensitivity and
collinearity indices used in environmental science and ecology \citep{Brun2001, Soetaert2010}, particularly if computational effort of MCMC is comparable to
estimation of profiles. If the collinearity index is above a heuristic threshold
for a given set of parameters then they may be considered practically
non-identifiable. Another approach to assess identifiability is to evaluate an
overlap statistic of the posterior and the prior, where a high degree of overlap
indicates that a parameter was not practically identifiable \citep{Rutter2009, Garrett2000}. We outline how to compute this statistic after obtaining the
posterior sample in \S \protect\hyperlink{overlap-statistic}{Overlap statistic}
(Supplementary Material 1).

In a calibration we must at least examine, if not formally establish, practical
identifiability for parameters with improper priors otherwise the posterior will
be improper \citep{Bayarri2004} and posterior sampling methods such as Markov Chain
Monte Carlo (MCMC) will not converge \citep{Raue2013}. If the model predictions are
insensitive to non-identifiable parameters, particularly those with
non-informative priors, then perhaps the model could be simplified \citep{Brun2001}
and the previous steps are revisited with the simpler model. Otherwise,
additional data either in the form of an informative prior or as part of the
observation process (likelihood) is necessary to proceed. A value-of-information
analysis may identify the observations that can provide the most additional
information for the parameter \citep{Jackson2019}.

\hypertarget{sample-from-the-posterior-distribution}{%
\subsection{Sample from the posterior distribution}\label{sample-from-the-posterior-distribution}}

Our next step is to obtain a sample from the posterior of the model's
parameters. MCMC methods are widely used for sampling from the posterior, in
particular the Metropolis/Metropolis-Hastings, Gibbs, and Hamiltonian Monte
Carlo samplers (references for specific algorithms provided below). These
algorithms use a Markov chain whose steady-state distribution is the posterior.
The sample is obtained by simulating a chain for long enough that convergence
can be reasonably established, and the ensuing elements of the simulated chain
form the sample. We outline the Metropolis algorithm in \S \protect\hyperlink{metropolis-sampler}{Metropolis
sampler} in Supplementary Material 1 as an example.

Running multiple chains can assist in monitoring convergence by considering the
between and within-chain variance \citep[see][\S6]{Brooks2011}. The ratio of these,
after discarding a fixed portion of the start of the chain, should be close to
one if the chain has converged. The converse statement does not hold, the chain
may not have converged even if the ratio is close to one. The tail of the
multiple chains, where they have converged, can then be pooled to form the
posterior sample.

The starting position of each chain should be drawn from a distribution that is
over-dispersed with respect to the posterior. This increases the likelihood that
the eventual sample is representative of the domain of the posterior, and not
only some neighbourhood of the starting positions. For problems where search
algorithms can find such estimates; a suitable initial state for the chain could
be drawn from the asymptotic distribution of the MLE (or similar for MAP
estimate), with a scaled covariance to increase dispersion. Informative prior
distributions may also be useful for drawing the initial state for each chain.

The effective sample size \citep[see][\S11.5 p286]{Gelman2013}, denoted
\(n_\text{eff}\), can be used to determine at what point to end the simulation of
(each) chain. The desired \(n_\text{eff}\) will depend on the precision required
for predictions and tests that will be performed with the model. For example,
the order of the error in the posterior mean of an approximately normally
distributed parameter is \(\sim 1 / \sqrt{n_\text{eff}}\), and the variance of the
posterior sample is dominated by the variance of the posterior itself, as
opposed to the uncertainty introduced via a finite sample thereof, with
\(n_\text{eff}\) as low as 100 \citep[see][\S10.5 p267]{Gelman2013}.

The algorithms can be tuned for better computational performance, although
they are all restricted by how efficiently the likelihood can be calculated.
Efficient implementations of the sampling algorithms are readily available, such
as stan \citep{Gelman2015}, NIMBLE \citep{DeValpine2017} (itself an extension of BUGS
\citep{Lunn2012}), and PyMC3 \citep{Salvatier2016}.

\hypertarget{assess-and-select-a-model}{%
\subsection{Assess and select a model}\label{assess-and-select-a-model}}

The next step is to measure the consistency of the model's predictions with the
calibration data. This is a test of \emph{internal validity} as described by
Dahabreh et al \citep{Dahabreh2017}. Models with insufficient consistency are
unlikely to make reliable predictions. Consistency can be measured using
goodness-of-fit or model discrepancy statistics, or information criteria. We
can often include or exclude effects or terms, or make alternative structural
choices, to create a collection of models that can be compared using these
measures. One or more of the best performing models would be selected from the
available models to be used for making predictions.

Alahmadi et al. \citet{Alahmadi2020} stated, ``The gold standard for selecting between
models while accounting for prior information is to either calculate Bayes
factors or the model evidence.'' However, Bayes factors can be computationally
intractable and non-informative priors can lead to ambiguity \citep{Berger2001}. As
an alternative, information criteria such as the Deviance Information Criterion
(DIC) and the Watanabe-Akaike Information Criterion \citep{Jackson2015} (and
\citep[see][\S7.2]{Gelman2013}) can be used. They are easily evaluated with a sample
from the parameter posterior and both criteria penalise models that may have
worse \emph{predictive} accuracy than a simpler model.

Neither the Bayes factor nor the information criteria are intuitive measures of
whether a model is sufficient for a particular application. The sum of square
error is easy to interpret and is a reasonable measure of performance when the
error is normally distributed with zero mean \citep[see][\S7.1 p167]{Gelman2013}.
However, most models are inadequate, giving rise to `model discrepancy' (or
`structural uncertainty') which is the difference between the mean model and
mean true output, given (true) input values \citep{Kennedy2001}. The discrepancy
function can be modelled and estimated, and summary measures thereof can
describe performance, guide model refinement and reduce uncertainty, or the
estimate may be included as a correction to the model \citep{OHagan2006}.

\hypertarget{use-the-posterior-sample}{%
\subsection{Use the posterior sample}\label{use-the-posterior-sample}}

Model calibration could be considered complete after the model selection step,
however the parameter-posterior sample can be used in ways that a single
estimate (or some other form of sample) from other procedures can not. The
salient usage is to calculate summary statistics from the sample to describe
uncertain knowledge of the parameters or predictions of the model.

We can describe the values of model-parameters compatible with the observed data
using the mean, standard deviation, or quantiles of the parameter-posterior
sample. Intervals such as the equal-tailed 95\% posterior credible interval,
which is estimated by the 2.5\textsuperscript{th} and 97.5\textsuperscript{th} percentiles of the sample,
describe the range of values of the parameter that would be unsurprising to
observe if we were able to measure the true value for the model.

Other quantities derived from the model can be described in a similar manner to
parameters. Expected values of quantities can be described by generating one
expected value for each sample of the parameter-posterior. Predictions follow
the same pattern, one sample from the predictive distribution of an observation
(or hypothetical observation) should be drawn for each parameter sample. The
posterior predictive p-value, a Bayesian `p-value', is defined as the
probability that the model prediction is more extreme than some (observed) value
\citep[see][\S6.3 p146]{Gelman2013}. Given a single draw, denoted
\(x_{\text{pred},i}\), from the predictive distribution of \(X\) for each of the \(n\)
samples from the parameter-posterior, the Bayesian p-value for one-sided test of
a statistic, denoted \(T(x,\theta)\), can be estimated by

\[
    \text{p-value} \approx \frac{ 1 }{ n }
        \sum_i \mathbb{I}_{ T(x_i, \theta_i) \geq T(x, \theta_i), }
\]

where \(\mathbb{I}\) denotes the indicator function.

\hypertarget{estimates-of-australian-smoking-behaviour-using-bayesian-calibration}{%
\section{Estimates of Australian smoking behaviour using Bayesian calibration}\label{estimates-of-australian-smoking-behaviour-using-bayesian-calibration}}

\hypertarget{introduction-1}{%
\subsection{Introduction}\label{introduction-1}}

Adult daily smoking rates in Australia have reduced from 25\% in 1991 to 11.6\%
in 2019 \citep{AIHW2020}. While quantifying the contribution of specific policy
interventions is difficult, evidence shows that the main drivers of prevalence
decline have been price control (tobacco taxes), hard-hitting mass media
antismoking campaigns and smoke-free public places \citep{Greenhalgh2020}. The
National Tobacco Strategy in 2012 set a policy target of 10\% adult daily smoking
prevalence by 2018 and an update on the strategy is yet to occur. A flexible and
dynamic model of tobacco smoking prevalence for the Australian population can
inform policy targets by quantifying the impact of ongoing and new tobacco
control strategies on future trends in smoking, burden of disease, and
cost-effectiveness evaluations. To this end, we built a model of the life-course
smoking behaviour of the Australian population. The model was similar to the
U.S. Smoking History Generator (SHG) developed by the Cancer Intervention and
Surveillance Modeling Network \citep{Holford2014A} that has been used to estimate the
impact of tobacco policy \citep{Tam2018} and the benefits, harms and
cost-effectiveness of lung cancer screening in the US \citep{Criss2019, tenHaaf2020}. During the model building process we sought to, within practical
limits, place calibration on firm statistical foundations to fulfil a decision
maker's need for dependable and meaningful measures of uncertainty
\citep{Briggs2012}.

We used a calibration methodology that incorporated identifiability analyses,
Bayesian evidence synthesis of multiple data sources, and uncertainty
quantification, as outlined in the first part of the manuscript. Through these
analyses we gained the following advantages over earlier smoking behaviour model
calibration efforts: 1) we could show that model predictions were measurably
improved by allowing those who had quit smoking long-term to switch to reporting
as having never smoked; 2) we quantified uncertainty in trends of smoking
initiation and cessation due to the random sampling in the surveys and trials
used to calibrate the model; and 3) we accounted for competing events exactly as
all transitions were calibrated together.

\hypertarget{methods}{%
\subsection{Methods}\label{methods}}

The model structure was an extension of that by Gartner, Barendregt and Hall
\citep{Gartner2009} who predicted smoking prevalence to the year 2056, using:
Australian survey data on smoking from 1980 to 2007, population counts and
mortality data, and smoking status-specific hazard ratios of death estimated
in the American Cancer Society's Cancer Prevention Study (CPS) II cohort. In
brief, Gartner et al.~calculated smoking prevalence by combining smoking-related
excess mortality with: the estimated sex-specific trend in the proportion that
ever-smoked in each cohort, and the age, sex, and year-specific annual
proportional change in smoking prevalence. They calibrated the model via a
weighted least-squares fit to the observed prevalence, and quantified
uncertainty in the predictions using a bootstrapping procedure.

In our model calibration we included additional surveys to extend the
observation period of smoking behaviour to the years 1962-2016 \citep{Vaneckova2021}.
We estimated smoking status-specific hazard ratios of death from the Sax
Institute's 45 and Up Study \citep{45AndUp2007} using methods published previously
\citep{Banks2015}. Although clear trends in smoking status-specific hazard ratios of
death have been observed over the last 50 years, primarily due to the relative
improvement in non-smoker mortality \citep{Thun2013}, estimates derived from the 45
and Up Study were more relevant to the Australian population and reflected the
smoking-related mortality for cohorts that made up much of the smoking survey
observations used to calibrate the model.

The annual proportional change in smoking prevalence estimated by Gartner et
al.\citep{Gartner2009} was a measurement of the effect of initiation, quitting and
relapse events together, conditional upon one-year survival. In our formulation,
we estimated the rate of quitting events alone, without needing to estimate
unsuccessful attempts and subsequent relapse, with the help of a `recent
quitter' state. We also estimated the rate ex-smokers switched to reporting as a
never smoker \citep{Kenkel2003, Visalpattanasin1987} by the inclusion of a state
within the model called `reporting as never'.

The structure of the Bayesian calibration procedure is outlined in the first
part of this manuscript (\S \protect\hyperlink{bayesian-calibration-of-simulation-models-tutorial}{Bayesian calibration of simulation models:
tutorial}). We started
with an analysis to determine which parameters in the model could be estimated
with ideal prevalence data, from which we determined that independent data on
the hazard ratio of death for either ex-smokers or smokers relative to never
smokers was necessary for successful calibration. The hazard ratio data were
then incorporated into calibration as part of a prior estimate of the parameter
values. We derived a mathematical expression to describe the survey sampling
process and identified a suitable parameter-value neighbourhood for a Markov
Chain Monte Carlo (MCMC) algorithm to sample from. We obtained 200 samples of
parameter values using the MCMC algorithm for each model under consideration. We
selected the best model out of eight and summarised trends in the life-course of
smoking behaviour using a sample of its; proportion that initiate, quit rate,
and rate that ex-smokers switch to reporting as never smokers.

\hypertarget{data-sources}{%
\subsubsection{Data sources}\label{data-sources}}

\hypertarget{smoking-status}{%
\paragraph{Smoking status}\label{smoking-status}}

We used individual-level data from 26 cross-sectional surveys conducted between
1962 and 2016, which were previously used to examine birth-cohort specific
trends in historical smoking prevalence \citep{Vaneckova2021}. The surveys provided
by the Australian Data Archive (ADA) and Cancer Council Victoria (CCV) were:

\begin{itemize}
\tightlist
\item
  National Drug Strategy Household Survey (NDSHS) 1998, 2001, 2004, 2007,
  2010, 2013 and 2016 \citep[see][references therein]{AIHW2017}.
\item
  National Campaign Against Drug Abuse and Social Issues Survey (NCADASIS)
  1991 and 1993 (including the Victorian Drug Household Survey 1993 sample),
  NDSHS 1995 (including the Victorian Drug Strategy Household Survey 1995
  sample), Social Issues Australia (SIA) survey 1985 \citep[see][
  references therein]{McAllister1996}.
\item
  Risk Factor Prevalence Study (RFPS) 1980, 1983 and 1989 \citep[see][
  references therein]{Risk1990}.
\item
  CCV Australian adult smoking surveys in 1974, 1976, 1980, 1983, 1986, 1989,
  1992 and 1995 \citep[see][references therein]{Hill1998}.
\item
  Australian Gallup Polls (AGP) no. 158, 160, 168 and 193 (1962-1967)
  \citep{Gallup1962, Gallup1963, Gallup1964, Gallup1968}.
\end{itemize}

We defined a current smoker as those who smoked on a daily/regular basis, with
the remainder of non-missing cases classified as non-smokers. In all surveys
except the Gallup Polls, non-smokers were further categorised as either
ex-smoker, recent quitter or never smoker. We defined ex-smokers as non-smokers
who had smoked on a daily/regular basis and had quit at least two years prior to
the survey, while those who had quit less than two years prior were recent
quitters. For surveys without a question on past daily smoking, or missing
responses, we used an affirmative response as to whether a participant had
smoked 100 cigarettes in their life to identify ex-smokers and recent quitters.
We defined all other non-smokers as never smokers. For participants with
sufficient information on a quit event, we further divided ex-smokers and recent
quitters into age-at-quit categories.

The surveys contained 254,231 observations in total. We
singly-imputed age in years for those whose age-at-survey was grouped using a
Penalised Composite Link Model fitted to each survey (i.e.~60,131 imputed ages) \citep{Rizzi2015}. We included participants of age
20-99 years at the time of
the survey, born from 1910 onwards (229,582 participants). We excluded 3,394
observations for missing smoking status, leaving a total of 226,188 observations. The rate of missing
information on whether a non-smoker, who had smoked in the past, had quit
within two years prior to the survey was less than 8.6\% in all but the 1985-95 NDSHS and the
1989 CCV survey. The 1985-95 NDSHS did not include the relevant questions, while
only a subset of ex-smokers in the 1989 CCV survey were asked how long ago they
had stopped. We defined all these missing cases as ex-smokers, as the clear
majority (86\%) of complete cases had quit at
least two years prior.

Non-random sampling occurred in the surveys, either due to a multi-stage
stratified or quota-sampling designs, or due to non-response bias. All surveys
supplied weights (except for the AGP in 1967, the CCV surveys in 1974 and 1980,
the NCADASIS in 1991 and the additional Victorian sample of the NDSHS in 1995)
which were designed to match the weighted count of responses to Australian
population data by age, sex and in some cases, other demographic variables. We
applied an iterative proportional fitting procedure \citep{Deville1993, Battaglia2009} to all surveys to estimate a correction to the supplied
weights such that the re-weighted response counts matched Australian Bureau of
Statistics population counts by age, sex, state, and capital city
versus non-capital city residence.

\hypertarget{mortality}{%
\paragraph{Mortality}\label{mortality}}

We obtained historical mortality data for the Australian population for the
period 1930-2016 from the Human Mortality Database: \href{https://mortality.org}{mortality.org} \citep{HMD2019}.

\hypertarget{hazard-ratios}{%
\paragraph{Hazard ratios}\label{hazard-ratios}}

We estimated smoking-status specific hazard ratios of death from the 45 and Up
Study, a prospective cohort study of 267,153 residents of NSW aged
$\geq$ 45 years at baseline (2006-2009), randomly sampled from the Services
Australia (formerly the Department of Human Services and Medicare Australia)
enrolment database \citep{45AndUp2007}. People 80+ years of age and residents of
rural and remote areas were oversampled by design, and the overall response rate
was about 18\% of invitees. The study oversampled higher income groups than the
target population \citep{Johar2012}, however hazard ratios derived from the study
may be generalisable to the population \citep{Mealing2010}.

Fact of death to August 2017 was ascertained from the NSW Registry of Births
Deaths and Marriages, with data linkage performed by the NSW Ministry of
Health's Centre for Health Record Linkage (CHeReL; \href{https://www.cherel.org.au}{www.cherel.org.au}). We applied Cox proportional hazards regressions separately by sex
for each 5-year age stratum from 45 to 99 years to estimate the hazard ratios of
death for current smokers, recent quitters, and ex-smokers compared to never
smokers using similar methods to those reported by Banks et al. \citep{Banks2015}
Briefly, we excluded participants for the following reasons: aged $\geq$ 100
years at baseline, data linkage errors, self-reported history of chronic disease
at baseline (heart disease, stroke, blood clot and cancer except melanoma and
non-melanoma skin cancer), and missing information on smoking status. Hazard
ratios were adjusted for age. After exclusions, 191,031 participants
were included for analysis. We assumed that current smokers and those who had
quit within two years (recent quitters) had the same hazard ratio. The analysis
was conducted within the Secure Unified Research Environment. Ethics approval
for the 45 and Up Study was provided by the University of NSW Human Research
Ethics Committee and approval for linkage by CHeReL was provided by the NSW
Population and Health Services Research Ethics Committee.

\hypertarget{model-structure}{%
\subsubsection{Model structure}\label{model-structure}}

\begin{figure}
\centering
\includegraphics{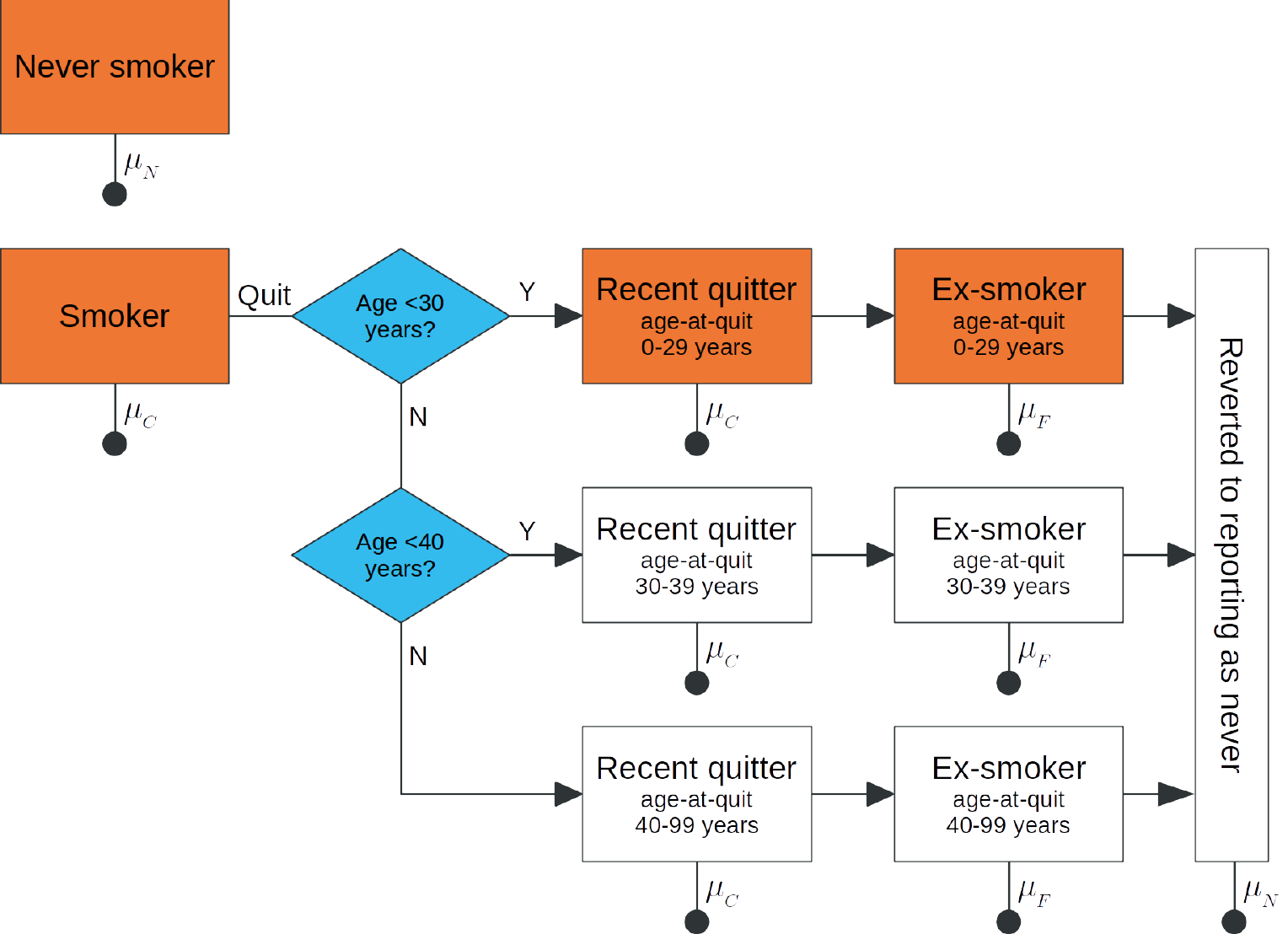}
\caption{\label{fig:structure-of-model}Structure of the compartmental model of lifetime smoking behaviour. Line with filled circle represents transition to death. Transition rates between boxes depend upon age and sex, except for the transition from recent quitter to ex-smoker which occurs at two years in the former state. \(\mu_N\) is the never-smoker mortality rate. \(\mu_C = HR_C \mu_N\) and \(\mu_F = HR_F \mu_N\), where \(HR_C\) and \(HR_F\) are the hazard ratios of death for current smokers and ex-smokers, respectively, compared to never smokers.}
\end{figure}

We used a compartmental model of the lifetime smoking behaviour of an
individual (or cohort), summarised in \textbf{Figure \ref{fig:structure-of-model}}.
The life-course of an individual proceeded in the following order:

\begin{enumerate}
\def\labelenumi{\arabic{enumi}.}
\tightlist
\item
  An individual either started smoking by a certain age (20 years), or they
  did not smoke throughout their lifetime.
\item
  If they started smoking;

  \begin{enumerate}
  \def\labelenumii{\arabic{enumii}.}
  \tightlist
  \item
    they would eventually quit and entered a `recent quitter' state,
    according to their age-at-quit group;
  \item
    after a fixed period (\(k=2\) years) they entered an `ex-smoker' state
    according to their age-at-quit group; and
  \item
    they could switch to reporting as a never smoker and thus enter the
    `reporting as never' state.
  \end{enumerate}
\end{enumerate}

We defined a quit event as the event that a smoker stops smoking and does not
smoke again. No explicit model of quit attempts and relapse was included. If
all relapse after a quit attempt occurred within \(k=2\) years in a surveyed
population then:

\begin{itemize}
\tightlist
\item
  The sum of current smokers and recent quitters in the model was consistent
  with the same sum in the surveys or the population (even if separately these
  terms were \emph{not}), and;
\item
  The estimate of the quit rate was consistent if; all smokers initiate before
  the selected age; migration is independent of smoking status; and the model
  of differential mortality was consistent with the population.
\end{itemize}

We assigned an age-at-quit category so that those who smoked for a shorter
duration may transition to reporting-as-never at a different rate from those
who smoked for longer. This was implemented as separate states for each
age-at-quit group within each smoking status that followed a quit event, as
shown in \textbf{Figure \ref{fig:structure-of-model}}.

At all times, an individual may die at a rate specific to their smoking status,
sex, and piece-wise constant by age:

\begin{itemize}
\tightlist
\item
  Never smokers and reporting-as-never smokers died at the same rate, denoted
  \(\mu_N\).
\item
  Current smokers and recent quitters died at the rate \(\mu_C = HR_C \mu_N\),
  with age-specific hazard ratio \(HR_C\).
\item
  Ex-smokers died at the rate \(\mu_F = HR_F \mu_N\) with age-specific hazard
  ratio \(HR_F\).
\end{itemize}

The initial population at age 20 years was made up of
never smokers, current smokers, and recent quitters or ex-smokers belonging to
the first age-at-quit group. The initial proportion of ex-smokers amongst
current smokers was the same for each cohort. The initial number of recent
quitters was determined using assumptions given in \S \protect\hyperlink{single-cohort-equations}{Single cohort
equations} (Supplementary Material 2).

We assumed the population mortality rate \(\mu\) (age, sex and year-specific)
satisfied

\[ \mu = \mu_N \rho_N + \mu_C \rho_C + \mu_F \rho_F, \]

where \(\rho_N\) was the total proportion in the never smoker and
reporting-as-never smoker categories, \(\rho_C\) was the total proportion in the
current smoker and recent-quitter categories, and \(\rho_F\) was the total
proportion in the ex-smoker categories.

\hypertarget{parameterisation}{%
\paragraph{Parameterisation}\label{parameterisation}}

The two key quantities for the initial population at age
20 years were the proportion of ex-smokers (amongst
current and ex-smokers), denoted \(P_F\), and the proportion that initiated,
denoted \(P_I\) (one minus the proportion of never smokers). We modelled the
logit-transformed proportion that initiated as a natural cubic spline-function
of the birth year.

We denote the quit rate as \(\lambda_Q\), and we modelled the log-transformed
value as the sum of natural cubic spline-functions of age and calendar year
(period). We assumed that the rate that ex-smokers transition to the
reporting-as-never state, denoted \(\lambda_R\), for each age-at-quit group (\textless30
years, 30-39 years, $\geq$40 years) was constant.

We pre-specified the knots for each spline term. Mathematical notation and
details of the knot locations considered are given in the \S \protect\hyperlink{spline-terms}{Spline
terms} in Supplementary Material 2. Along with different numbers
of knots, terms were included or excluded to form eight different models,
labelled `null' for the model with only intercept terms (and none reporting as
never) and then labelled `A' through `G', listed in \textbf{Table
\ref{tab:model-specification-summary}} in Supplementary Material 2. We assumed
that the total number of degrees of freedom, including the age-group specific
hazard ratios, will be small compared to the number of observations in the
surveys so that the parameters might be practically identifiable in each model.

\hypertarget{bayesian-calibration}{%
\subsubsection{Bayesian calibration}\label{bayesian-calibration}}

\hypertarget{parameters-to-calibrate}{%
\paragraph{Parameters to calibrate}\label{parameters-to-calibrate}}

To select suitable parameters to calibrate we informally examined structural
identifiability. We considered a simplified scenario where the proportions of
never, current and former smokers were known for a cohort. We argued, with
details provided in \S \protect\hyperlink{structural-identifiability-of-the-smoking-behaviour-model}{Structural identifiability of the smoking
behaviour model}
(Supplementary Material 2), that the quantities \(P_F\) and \(P_I\) could be
identified from the observed proportions at the starting age in the model. The
remaining four quantities, \(\lambda_Q\), \(\lambda_R\), \(\mu_C\) and \(\mu_F\), were
not identifiable, but if any one of these were known, the remaining three would
be identifiable. Therefore, if all of these parameters were included in the
calibration, we required at least one to have an informative prior, rather than
a non-informative prior.

\hypertarget{prior-distribution}{%
\paragraph{Prior distribution}\label{prior-distribution}}

We assumed that the joint prior of the parameters could be decomposed into
independent priors for; the hazard ratios, the proportion \(P_F\), the
coefficients of the spline describing \(P_I\) (including an intercept), the
coefficients of the sum of the splines in \(\lambda_Q\) (including an intercept),
and each rate of switching to reporting as never.

We used an informative prior for the hazard ratios, given by the asymptotic
distribution of the estimate from the Cox regressions. In doing so, we addressed
the non-identifiability described earlier by suppling information on both
\(\mu_C\) and \(\mu_F\). Although our examination of structural identifiability
suggested we only needed prior information on one of these two, there was no
reason to leave additional readily available information out of the calibration
process.

The proportion \(P_F\) was given a uniform prior on \([0, 0.5]\), indicating
indifference to any value in that range. We assigned data-dependent multivariate
normal priors to the coefficients of \(P_I\) and \(\lambda_Q\), detailed in \S
\protect\hyperlink{detailed-prior-distribution}{Detailed prior distribution} (Supplementary
Material 2), centered at no effect and intended to have minimal impact on
sampling parameter values from the posterior compared to the impact of the
observed data. One nuisance parameter was introduced for each data-dependent
prior. We assigned each rate of switching to reporting as never, \(\lambda_R\),
the prior \(1 / \sqrt{\lambda_R}\).

\hypertarget{model-of-smoking-survey-responses}{%
\paragraph{Model of smoking survey responses}\label{model-of-smoking-survey-responses}}

The dependence structure of the priors, the model, the smoking survey data,
and the population mortality data is shown in \textbf{Figure
\ref{fig:calibration-graphical-model}} (Supplementary Material 2). In order to
sample parameter values within this structure we required an expression for
likelihood of the smoking survey data.

We assumed each survey response was drawn from a categorical distribution, with
sex-, age-, and birth year-specific probabilities. The three categories were
named `current', `ex-' (by age-at-quit group), and `never' smokers, with
definitions as per above in \S \protect\hyperlink{data-sources}{Data sources}, except for
the recent-quitter category being merged into the current-smoker category. We
assumed the probabilities for each category were the same as the sum of one or
more expected proportions in the model. The current-smoker probabilities were
given by the sum of the model's current smokers and recent-quitter proportions,
ex-smoker probabilities were given by the model's ex-smoker proportions, and
never-smoker probabilities would be the remainder. We found the expected
proportions by numerically solving the governing equations of the model,
detailed in \S \protect\hyperlink{single-cohort-equations}{Single cohort equations}
(Supplementary Material 2).

We accounted for uncertainty introduced by non-random sampling using the
effective sample size of each cell in the cross-tabulation of each survey by
sex, age and birth year. The likelihood for each independent cell was given by a
Dirichlet distribution with parameters determined by observed weighted
proportions and the effective sample size, details provided in \S
\protect\hyperlink{survey-data-likelihood}{Survey data likelihood} (Supplementary Material 2).

\hypertarget{suitable-sampling-region}{%
\paragraph{Suitable sampling region}\label{suitable-sampling-region}}

We found a suitable region of parameter values to sample from by investigating
local practical identifiability of the maximum a priori (MAP) estimate of the
parameters. For this step, only, we fixed the hazard ratios at their prior mode.
We defined local practical identifiability as the posterior-based 95\%
`confidence' interval (CI) of the estimate being finite in extent \citep{Raue2013}.
We used the \texttt{optim()} command in R \citep{R2020} to find the MAP estimate. We
calculated the profile posterior of each component using the predictor-corrector
approach defined in the \S \protect\hyperlink{numerical-methods-for-practical-identifiability}{Numerical methods for practical
identifiability}
(Supplementary Material 1), in the neighbourhood given by \(\pm\) 7.1 standard deviations of the asymptotic
distribution of the MAP estimate. We then estimated the highest level of
confidence for which the CI was contained within the neighbourhood; if the level
was less than 95\% then practical identifiability of the parameter near the MAP
estimate was questionable.

The degree to which the hazard ratios were identified by the survey data was
investigated with the overlap statistic \citep{Garrett2000, Rutter2009} which was
calculated using the approach outlined in \S \protect\hyperlink{overlap-statistic}{Numerical methods for
practical identifiability} (Supplementary Material 1) and
the posterior sample obtained in the next step; the threshold 0.35 or less was
used as evidence of weak identifiability.

\hypertarget{parameter-value-samples}{%
\paragraph{Parameter-value samples}\label{parameter-value-samples}}

We used the Metropolis-within-Gibbs algorithm to sample from the joint posterior
of the hazard ratios, the model parameters, and the nuisance parameters. Each
of these was the basis of a block in the Gibbs sampler. Five chains were simulated, starting
at random points drawn from a distribution overdispersed with respect to the
asymptotic distribution of the maximum likelihood estimate (MLE) of the hazard
ratios and the MAP estimate of the other parameters. After a burn-in phase of
length 1600 samples per chain, the
samples were discarded and the algorithm resumed until an estimated effective
sample size of 50 was obtained \citep[see][equation 11.8]{Gelman2013}. We
culled each chain to 40 evenly-spaced
samples, for a total of 200 samples of parameter values. Further
details of the method used are supplied in \S \protect\hyperlink{metropolis-within-gibbs-sampler}{Metropolis-within-Gibbs
Sampler} in Supplementary Material 2. We used
this procedure for each model listed in \textbf{Table
\ref{tab:model-specification-summary}} (Supplementary Material 2).

\hypertarget{model-selection}{%
\paragraph{Model selection}\label{model-selection}}

We estimated the Deviance Information Criterion (DIC) for each model to assess
improvement in predictive accuracy between models. The procedure to calculate
the DIC using the samples from the joint posterior is detailed in \S
\protect\hyperlink{deviance-information-criterion}{Deviance information criterion} (Supplementary
Material 2). We applied some discretion in selecting a model if improvement in
the DIC over a nested model was small, to reduce risk of over-fitting.

We summarised the model discrepancy in the proportion of smokers, the proportion
of never smokers amongst non-smokers, and the proportion of ex-smokers that quit
before age 30 years. The discrepancy was assumed to be a Gaussian process
\citep{Kammann2003} and was estimated by regression of the residual proportions (the
difference between the model-expected and survey values) using the \texttt{mgcv}
package in R for Generalised Additive Models \citep{Wood2006, Wood2017}. The
discrepancy was summarised using the expected mean and standard deviation on a
log-odds scale of the discrepancy for the observed survey data; with the
expectation taken over the parameter-posterior.

\hypertarget{cross-sectional-smoking-prevalence}{%
\paragraph{Cross-sectional smoking prevalence}\label{cross-sectional-smoking-prevalence}}

We generated 200 samples of the predicted proportions of never,
current, and ex-smokers for each survey by sex and calendar year. Each sample
corresponded to one prediction from the model using one of the samples of the
parameter values. We compared the observed proportions in the NDSHS 2016 to
model predictions via the probabilities that each model prediction was more
extreme than the observed value. The probabilities, which are also known as
Bayesian posterior predictive p-values, were estimated using the corresponding
proportions of more extreme values in the sample. We modelled the predicted
counts, used to determine the predicted proportions, as Dirichlet-Multinomial
random variables as described in \S \protect\hyperlink{survey-data-likelihood}{Survey data likelihood} in Supplementary Material 2.

\hypertarget{trends-in-life-course-of-smoking}{%
\paragraph{Trends in life-course of smoking}\label{trends-in-life-course-of-smoking}}

We generated 200 samples of;

\begin{itemize}
\tightlist
\item
  the expected proportion that initiated before age
  20 years by birth year and sex;
\item
  the expected rate of quitting smoking by age, calendar year, and sex; and
\item
  the expected rate that ex-smokers would switch to reporting as never in a
  survey by sex and age-at-quit group.
\end{itemize}

Each sample corresponded to the expected value of the model using one of the
samples of the parameter values. We examined evidence for differences between
men and women using the expected probability that the value for men was more
extreme than the value for women. We estimated the probabilities using the mean
proportion of more extreme values in the men's sample compared to each value
from the women's sample.

The quit rate cannot be validated directly using the survey data as the quit
event in our model is conditional upon no future relapse and we do not know
which non-smokers in the survey will relapse. However, in the ITC Four Country
Survey, a longitudinal study, it was observed that 95\% of those with at least
two years abstinence went on to maintain abstinence over the next year
\citep{Herd2009}, therefore we compared the model to an estimate of the quit rate
from two years prior to a survey given sustained abstinence of at least two
years.

\hypertarget{sensitivity-analyses}{%
\subsubsection{Sensitivity analyses}\label{sensitivity-analyses}}

We tested the sensitivity of the model predictions to different choices in the
calibration of the selected model. We tested the choices of: 1) the prior for
the hazard ratios of death in relation to smoking status; 2) the age at which
initiation of smoking was completed within a cohort; 3) using a cohort term in
the quit rate as opposed to a calendar year term; 4) whether those who quit
after age 40 years could report as a never smoker; 5) allowing any ex-smoker to
report as never; and 6) whether survey weights were used in the likelihood. In
each sensitivity analysis we modified the selected model and obtained a sample
from the parameter posterior, estimated the DIC for all but item 2) in the
preceding list (as it was not comparable), and we generated expected values of;
the sex-specific proportion that initiated by age 20 years in the latest cohort
(born 1996), the age and sex-specific rate of quitting daily smoking
in the year 2016, and the sex and age-at-quit group-specific rate that
ex-smokers switch to reporting as never.

Contemporary estimates of the hazards of smoking-related mortality may not be
valid throughout the observation period, particularly for earlier years when the
hazard ratios are known to be lower due to changes in smoking intensity, age at
initiation, and never-smoker mortality \citep{Thun2013}. To assess the sensitivity of
the model predictions to the selection of prior for the hazard ratios, we
calibrated the selected model using estimates of hazard ratios derived from the
12-year follow-up of the CPS-I cohort \citep{Burns1997}, recruited in 1959-1960, in
the prior instead of the 45 and Up Study estimates. Given that the estimates of
the hazard ratios have increased over time in the US \citep{Thun2013} this was a
prior that greatly under-estimated smoking-related mortality in the later period
of the smoking survey data.

In the survey data, only about 70-85\% of smokers within any birth cohort had
started smoking by age 20 years. A more realistic age at completed initiation
would be 25 years by which age 93-97\% of smokers had started \citep{Vaneckova2021}.
To assess the impact of assuming all smokers start by age 20 years, we
calibrated the selected model using a starting age of 25 years.

For simplicity (and identifiability) we chose that the quit rate was a sum of
age and calendar year terms, which imply at most a linear relationship with
cohort. We considered an alternative in sensitivity analysis by using a sum of
age and cohort terms, implying at most a linear calendar year effect. The DIC
can be used to determine which choice is more compatible with the survey data,
i.e. which non-linear effect out of calendar year and cohort is more
useful.

To assess the sensitivity of the model to the assumption that those who smoked
and quit after the age of 40 years would always report as an ex-smoker, and not
switch to reporting as a never smoker, we calibrated the selected model with
the assumption of a non-zero rate of switching for the later age-at-quit
group. Similarly, we tested no reporting as a never smoker for all age-at-quit
groups if the selected model did allow switching.

In the main analysis we used the approximate effective sample size of the
smoking survey data and individual weights to account for survey design effects.
To determine how sensitive our results were to the adjustment for non-random
sampling, we also calibrated the selected model ignoring the weights and
effective sample size in the likelihood.

\hypertarget{results}{%
\subsection{Results}\label{results}}

\hypertarget{calibration}{%
\subsubsection{Calibration}\label{calibration}}

The neighbourhood of the MAP estimate for each model met our requirements for a
suitable region for sampling parameter-values. The Hessian of the posterior
function at the MAP estimate was positive definite for each model, therefore the
neighbourhoods all contained a local maximum. Practical identifiability in the
neighbourhood was supported by the confidence-level test in all but a few cases.
Notable exceptions were the nuisance parameters, for which the highest
confidence level with associated interval contained within the (compuuted)
neighbourhood varied between 65\%
and 98\% depending on the model.
For the parameter describing the rate that male ex-smokers switched to reporting
as a never smoker, with age-at-quit 30-39 years, the level was no greater than
63\% in any model. The
highest confidence level for which the corresponding interval was contained
within the computed neighbourhood is provided for each model in \textbf{Table
\ref{tab:profile-posterior-confidence-summary}} (Supplementary Material 2).
The hazard ratios of mortality did not meet the criteria of weak identifiability
according to the overlap statistic; the distributions of the prior hazard ratios
of mortality and the posterior hazard ratios are summarised by their median and
5\textsuperscript{th} and 95\textsuperscript{th} percentiles in \textbf{Table \ref{tab:hazard-ratio-overlap}}
(Supplementary Material 2), along with the overlap statistics.

The parameter-value samples from the MCMC algorithm satisfied our criteria
regarding (non-)convergence. All samples reached an effective sample size of at
least 50.0, and the estimated
potential variance reduction was at most 1.11 for any component after 2472 iterations per chain.

The DIC estimate for model `F' improved upon all simpler models, and model `G'
only demonstrated a marginal improvement for men, as shown in \textbf{Table
\ref{tab:model-deviance-information-criterion}} (Supplementary Material 2). To
minimise risk of over-fitting, we selected model `F' for the main analysis.
Model `F' had: three degrees of freedom (d.f.) in the effect of birth year in
the proportion of a cohort that initiated; two d.f. each for the effects of age
and calendar-year in the quit rate; and non-zero rates of switching to reporting
as never by age-at-quit group for those who quit before age 40 years.

The measures of discrepancy of the selected model were smaller than those for
the `null' model. The magnitude of the mean discrepancy - a measure of bias -
in either the proportions of smokers in the population or the proportions of
never-smokers amongst non-smokers, both on a log-odds scale, was at most
0.023 in the selected model, compared to at most
0.170 for the `null' model. Likewise, the standard
deviation - a measure of the unexplained variation - was no greater than
0.153 in the selected model, compared to at most
0.498 in the `null' model. The magnitudes of the mean and the
standard deviations of the discrepancy in the proportion of ex-smokers that had
quit prior to age 30 years were greater than for the other proportions (for a
given model), with values for the selected model at most
0.119 and 0.405 respectively. The
mean and standard deviation in the discrepancy for each model and each
proportion evaluated is shown in \textbf{Table \ref{tab:discrepancy-in-proportions}}
(Supplementary Material 2).

\hypertarget{model-estimates}{%
\subsubsection{Model estimates}\label{model-estimates}}

\hypertarget{cross-sectional-smoking-prevalence-1}{%
\paragraph{Cross-sectional smoking prevalence}\label{cross-sectional-smoking-prevalence-1}}

The predicted cross-sectional proportion of Australian men and women age
$\geq$20 years in each smoking status category is shown, by survey, in
\textbf{Figure \ref{fig:cross-section-smoking-status}}, comparing the survey point
estimate with the interval given by the 5\textsuperscript{th} and 95\textsuperscript{th} percentiles
of predictions sampled from the model (the 90\% equal-tailed intervals or ``90\%
ETI'' hereafter). 52.0\% of the survey
proportions were within the intervals, considerably less than the ideal
value of 90\%. The Pearson correlations between the
survey data and the model-sampled predictions are displayed on each panel by
sex and smoking status in \textbf{Figure \ref{fig:cross-section-smoking-status}}.
The correlations were lower for the quantities with more non-linear calendar
year trends, e.g.~90\% of the correlations for the
proportion of never smokers amongst women were in the interval {[}0.810,0.900{]}, and were higher for simpler
trends, e.g.~90\% of the correlations for the
proportion of current smokers amongst men were in the interval {[}0.969,0.982{]}.

\afterpage{
\begin{landscape}
\thispagestyle{lscape}
\pagestyle{lscape}
\begin{figure}[t]
\centering
\includegraphics{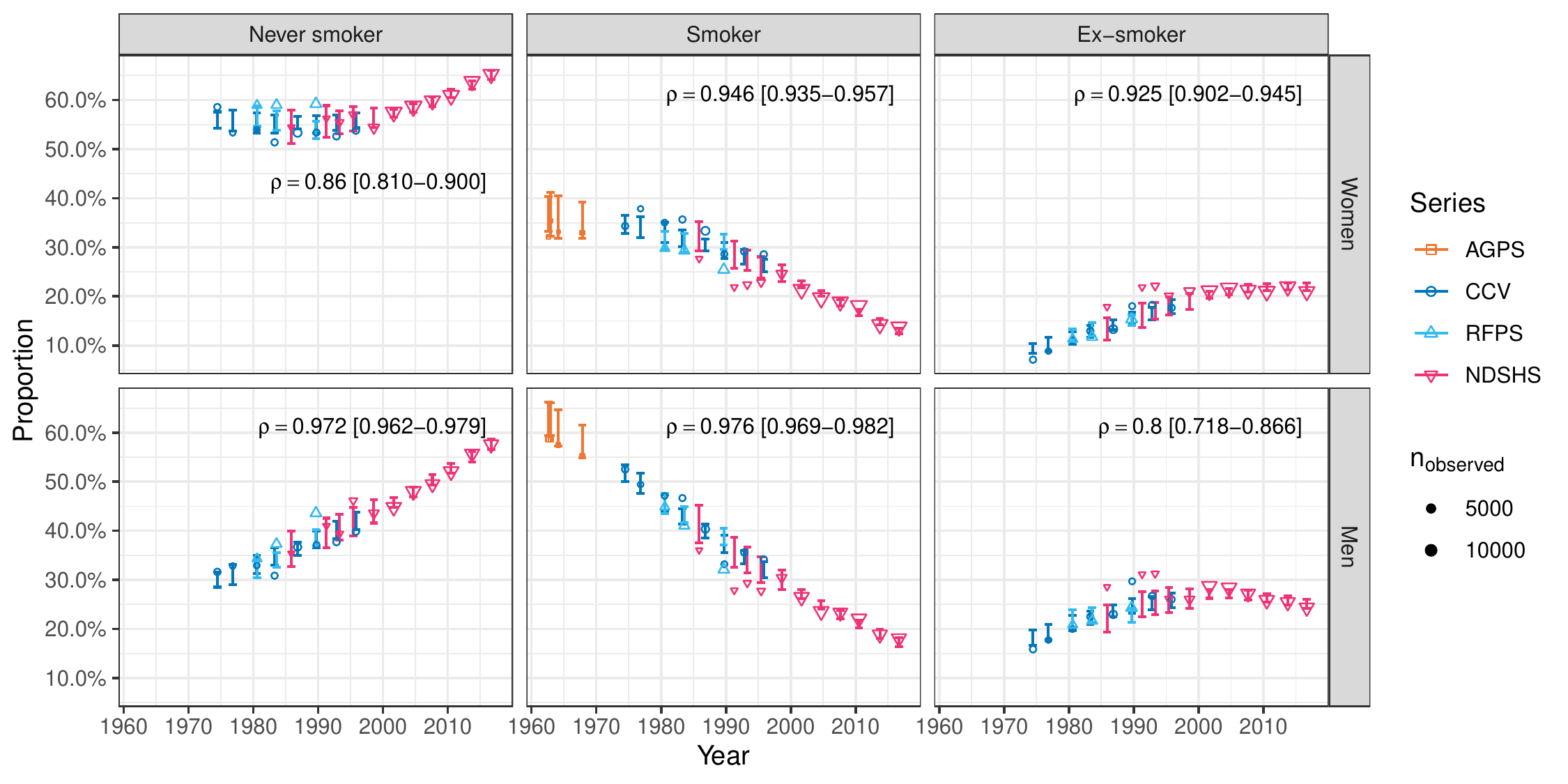}
\caption{\label{fig:cross-section-smoking-status}Proportion of never smokers, smokers and ex-smokers amongst Australian men and women age $\geq$20 years, in four smoking survey series conducted between 1962-2016 (markers), and corresponding intervals given by the 5\textsuperscript{th} and 95\textsuperscript{th} percentiles of the predictions sampled from the selected model (error bars). Size of marker indicates number of respondents age $\geq$20 years in each survey. Median Pearson correlation between survey data and the model-samples (and interval given by the 5\textsuperscript{th} and 95\textsuperscript{th} percentiles) shown on each panel.Abbreviations; AGPS: Australian Gallup Poll series; CCV: Cancer Council Victoria adult smoking surveys; RFPS: Risk Factor Prevalence Study; NDSHS: National Drug Strategy Household Survey.}
\end{figure}
\end{landscape}
}

The proportions of never smokers, smokers and ex-smokers amongst men, women,
and persons in the NDSHS in 2016 and the corresponding ETIs containing 90\% of
the sampled predictions from the model are shown in \textbf{Table
\ref{tab:cross-section-latest-NDSHS}}, along with the estimate of the p-value
that the predictions were more extreme than the survey value. The p-values for
the never-smoker proportions were all greater than 0.810, on the other hand, the p-values for the
proportions of smokers and ex-smokers were smaller, near to
0.180 amongst the predictions for men, and no greater than
0.030 for women or persons.

\afterpage{
\begin{table}[t]
\centering
\caption{\label{tab:cross-section-latest-NDSHS}Proportions of never smokers, smokers and ex-smokers amongst Australians $\geq$20 years of age in 2016 observed in the National Drug Strategy Household Survey, and the sample of prediction thereof obtained from the selected model, for men, women and persons. Predictions are described by the median and the interval given by the 5\textsuperscript{th} and 95\textsuperscript{th} percentiles of their samples. The p-value is the proportion of the model-sampled predictions that were more extreme than the observed value in the survey.}
\begin{tabular}{>{\bfseries}llrrr}
\toprule
Sex & \bfseries Status & \bfseries Model sample & \bfseries Observed & \bfseries p-value \\
\midrule
Women & Never smoker & 65.2\% {[}64.3\%,66.0\%{]} & 65.2\% & 0.840 \\
      & Smoker & 13.0\% {[}12.3\%,13.6\%{]} & 13.8\% & 0.030 \\
      & Ex-smoker & 21.9\% {[}21.1\%,22.7\%{]} & 21.0\% & 0.030 \\
Men & Never smoker & 57.5\% {[}56.6\%,58.7\%{]} & 57.6\% & 0.880 \\
    & Smoker & 17.3\% {[}16.4\%,18.2\%{]} & 18.1\% & 0.160 \\
    & Ex-smoker & 25.3\% {[}24.2\%,26.0\%{]} & 24.3\% & 0.180 \\
Persons & Never smoker & 61.4\% {[}60.7\%,62.1\%{]} & 61.5\% & 0.810 \\
        & Smoker & 15.1\% {[}14.5\%,15.6\%{]} & 15.9\% & 0.020 \\
        & Ex-smoker & 23.5\% {[}22.9\%,24.2\%{]} & 22.6\% & 0.020 \\
\bottomrule
\end{tabular}
\end{table}
}

\hypertarget{trends-in-initiation-cessation-and-reporting-smoking-status}{%
\paragraph{Trends in initiation, cessation, and reporting smoking status}\label{trends-in-initiation-cessation-and-reporting-smoking-status}}

The sample obtained from the model of the proportion that initiated daily
smoking in a birth cohort is summarised in \textbf{Table
\ref{tab:calibrated-ever-proportion}}, by sex, and in ten-year increments of
birth year 1910 to
1990 and birth year
1996. The proportion of women within a birth
cohort who initiated peaked for those born in 1962, with sample median being
55.3\% (90\% ETI: {[}54.6\%,55.9\%{]}). For women born in 1910, the sample median of the proportion was 44.2\% (90\% ETI: {[}42.3\%,46.1\%{]}). Since the peak, the values
decreased, and for women born in 1996, the last
cohort in the calibration period, the sample median was 16.3\% (90\% ETI: {[}15.2\%,17.5\%{]}). The proportion of men that
initiated smoking decreased with each birth year from 1910, for which the sample median of the proportions was
87.1\% (90\% ETI: {[}85.9\%,88.4\%{]}), to
1996, with median 22.7\% (90\% ETI: {[}21.1\%,24.2\%{]}). The mean proportion of
sampled values for men that were more extreme than that of women was less than
5\% up until birth year 1967, and again from 1981. The growth rate of the proportion that initiated
is summarised in \textbf{Table \ref{tab:calibrated-change-in-ever-smoker}} 9
(Supplementary Material 2), by sex, and by the same birth years as \textbf{Table
\ref{tab:calibrated-ever-proportion}}.

The sample obtained from the model of the quit rate per 100 person-years (PY)
amongst Australian daily smokers is summarised in \textbf{Table
\ref{tab:calibrated-quit-rates-full}} by sex, age (at 30, 50, and 70 years),
and ten-year increments of calendar year from 1940 to 2010 and
the calendar year 2016. The sampled quit rates increased with calendar
year throughout, while the effect of age varied by sex. In a given calendar year
women quit at higher rates at earlier and later ages compared to age 50 years.
In the last year of the calibration period, 2016, the median sampled
quit rate per 100PY for women at age 30 and 70 years was
5.52 (90\% ETI: {[}5.19,5.88{]})
and 6.05 (90\% ETI: {[}5.36,6.72{]}), respectively, whereas at age 50 years, the sample median was
5.00 (90\% ETI: {[}4.77,5.31{]}) events per 100PY. In a fixed calendar year, the quit rate for men
increased with age. In the year 2016 the median sampled quit rate per
100PY for men at age 30 years was 3.67 (90\% ETI: {[}3.44,3.92{]}), whereas at age 70 years the
sample median was 4.70 (90\% ETI: {[}4.13,5.29{]}). In 2016, the mean proportion of sampled values for men that
were more extreme than that of women was less than 5\% up until age
47 years, and again from age
65 years. The sample obtained for
the growth (rate) of the quit rate is summarised in \textbf{Table
\ref{tab:calibrated-quit-rate-growth}} (Supplementary Material 2), by sex, and
calendar year for the same years as \textbf{Table
\ref{tab:calibrated-quit-rates-full}}. The quit rate for women grew faster
than for men through the years 1930 until 2016.

Female ex-smokers switched to reporting as a never smoker at a higher rate than
male ex-smokers in both age-at-quit groups. The median sampled rate for men who
quit before age 30 years was 2.07 (90\% ETI: {[}1.94,2.23{]}) events per 100PY, and for
those who quit between ages 30 and 39 years it was 0.31 (90\% ETI: {[}0.15,0.45{]})
events per 100PY. For women, the median sampled rate for those who quit before
age 30 years was 2.31 (90\% ETI: {[}2.14,2.46{]}) events per 100PY, and for those
who quit between ages 30 and 39 years it was 0.86 (90\% ETI: {[}0.66,1.03{]}) events per
100PY. The mean proportion of more extreme values for men compared to women was
at most 0.063.

The sample of the proportion of those who started smoking before age 20 years
who had also quit smoking by age 20 years is shown, by sex, in \textbf{Table
\ref{tab:calibrated-quit-before-twenty}} (Supplementary Material 2).

The comparison of the estimate of the quit rate from two years prior to a survey
(given sustained abstinence of at least two years) to the model's quit rate
is shown in \textbf{Figure \ref{fig:cross-section-quit-rate}} (Supplementary
Material 2). The survey estimate is, on average, marginally higher than the
model, which may be partly explained by relapse after two years amongst some
non-smokers.

\begin{table}[ht]
\centering
\caption{\label{tab:calibrated-ever-proportion}Sample of the proportion of Australians that initiated daily smoking obtained from the selected model, by sex (men, women, and persons) and ten-year increments of birth year from 1910 to 1990 and birth year 1996. Shown are the sample median and the interval given by the 5\textsuperscript{th} and 95\textsuperscript{th} percentiles of each sample. p-value: M $\equiv$ W; the mean proportion of more extreme values for men compared to each value for women.}
\begin{tabular}{
  >{\raggedright\arraybackslash}p{(\columnwidth - 10\tabcolsep) * \real{0.14}}
  >{\raggedleft\arraybackslash}p{(\columnwidth - 10\tabcolsep) * \real{0.18}}
  >{\raggedleft\arraybackslash}p{(\columnwidth - 10\tabcolsep) * \real{0.18}}
  >{\raggedleft\arraybackslash}p{(\columnwidth - 10\tabcolsep) * \real{0.18}}
  >{\raggedleft\arraybackslash}p{(\columnwidth - 10\tabcolsep) * \real{0.15}}}
\toprule
\bfseries Birth year & \bfseries Women & \bfseries Men & \bfseries Persons & \bfseries p-value: M $\equiv$ W \\
\midrule
1910 & 44.2\% {[}42.3\%,46.1\%{]} & 87.1\% {[}85.9\%,88.4\%{]} & 66.4\% {[}65.6\%,67.5\%{]} & $\leq$0.001 \\
1920 & 43.3\% {[}42.1\%,44.3\%{]} & 82.3\% {[}81.2\%,83.4\%{]} & 63.3\% {[}62.6\%,64.0\%{]} & $\leq$0.001 \\
1930 & 43.7\% {[}42.9\%,44.4\%{]} & 76.8\% {[}75.8\%,77.7\%{]} & 60.5\% {[}60.1\%,61.1\%{]} & $\leq$0.001 \\
1940 & 46.4\% {[}45.8\%,47.2\%{]} & 71.1\% {[}70.3\%,71.9\%{]} & 59.1\% {[}58.6\%,59.7\%{]} & $\leq$0.001 \\
1950 & 51.7\% {[}51.1\%,52.4\%{]} & 65.9\% {[}65.2\%,66.6\%{]} & 58.9\% {[}58.4\%,59.4\%{]} & $\leq$0.001 \\
1960 & 55.2\% {[}54.5\%,55.8\%{]} & 60.1\% {[}59.4\%,60.7\%{]} & 57.7\% {[}57.2\%,58.1\%{]} & $\leq$0.001 \\
1970 & 52.1\% {[}51.5\%,52.8\%{]} & 51.9\% {[}51.3\%,52.6\%{]} & 52.0\% {[}51.6\%,52.5\%{]} & 0.830 \\
1980 & 40.0\% {[}39.3\%,40.7\%{]} & 40.9\% {[}40.1\%,41.6\%{]} & 40.5\% {[}40.1\%,40.9\%{]} & 0.154 \\
1990 & 24.3\% {[}23.3\%,25.4\%{]} & 29.0\% {[}27.8\%,30.3\%{]} & 26.8\% {[}25.9\%,27.5\%{]} & $\leq$0.001 \\
1996 & 16.3\% {[}15.2\%,17.5\%{]} & 22.7\% {[}21.1\%,24.2\%{]} & 19.6\% {[}18.6\%,20.4\%{]} & $\leq$0.001 \\
\bottomrule
\end{tabular}
\end{table}

\afterpage{
\clearpage
\begin{landscape}
\thispagestyle{lscape}
\pagestyle{lscape}
\begin{table}
\addtolength{\tabcolsep}{-2pt}
\centering
\caption{\label{tab:calibrated-quit-rates-full}Sample of the rate that Australian smokers quit daily smoking (per 100 person-years) obtained from the selected model by sex, selected ages (30, 50 and 70 years), and ten-year increments of calendar year from 1940 to 2010 and in 2016. Shown are the sample median and the interval given by the 5\textsuperscript{th} and 95\textsuperscript{th} percentiles of each sample. p-value: M $\equiv$ W; the mean proportion of more extreme values for men compared to each value for women.}
\begin{tabular}{
  >{\raggedright\arraybackslash}p{(\columnwidth - 20\tabcolsep) * \real{0.05}}
  >{\raggedleft\arraybackslash}p{(\columnwidth - 20\tabcolsep) * \real{0.11}}
  >{\raggedleft\arraybackslash}p{(\columnwidth - 20\tabcolsep) * \real{0.11}}
  >{\raggedleft\arraybackslash}p{(\columnwidth - 20\tabcolsep) * \real{0.10}}
  >{\raggedleft\arraybackslash}p{(\columnwidth - 20\tabcolsep) * \real{0.11}}
  >{\raggedleft\arraybackslash}p{(\columnwidth - 20\tabcolsep) * \real{0.10}}
  >{\raggedleft\arraybackslash}p{(\columnwidth - 20\tabcolsep) * \real{0.10}}
  >{\raggedleft\arraybackslash}p{(\columnwidth - 20\tabcolsep) * \real{0.11}}
  >{\raggedleft\arraybackslash}p{(\columnwidth - 20\tabcolsep) * \real{0.11}}
  >{\raggedleft\arraybackslash}p{(\columnwidth - 20\tabcolsep) * \real{0.10}}}
\toprule
  &
  \multicolumn{3}{c}{\bfseries Age 30 years} &
  \multicolumn{3}{c}{\bfseries Age 50 years} &
  \multicolumn{3}{c}{\bfseries Age 70 years} \\
                  \cmidrule(lr){2-4} \cmidrule(lr){5-7} \cmidrule(l){8-10}
\bfseries Year &
  \bfseries Women & \bfseries Men & \bfseries p-value: M $\equiv$ W &
  \bfseries Women & \bfseries Men & \bfseries p-value: M $\equiv$ W &
  \bfseries Women & \bfseries Men & \bfseries p-value: M $\equiv$ W \\
\cmidrule(r){1-1} \cmidrule(lr){2-4} \cmidrule(lr){5-7} \cmidrule(l){8-10}
1940 & 0.35 {[}0.28,0.42{]} & 0.50 {[}0.43,0.57{]} & 0.025 & - & - & - & - & - & - \\
1950 & 0.63 {[}0.54,0.72{]} & 0.83 {[}0.75,0.90{]} & 0.015 & - & - & - & - & - & - \\
1960 & 1.10 {[}1.00,1.18{]} & 1.29 {[}1.21,1.36{]} & 0.011 & 0.99 {[}0.92,1.07{]} & 1.65 {[}1.58,1.73{]} & $\leq$0.001 & - & - & - \\
1970 & 1.76 {[}1.67,1.84{]} & 1.87 {[}1.79,1.93{]} & 0.103 & 1.60 {[}1.52,1.66{]} & 2.40 {[}2.34,2.47{]} & $\leq$0.001 & - & - & - \\
1980 & 2.55 {[}2.46,2.63{]} & 2.45 {[}2.37,2.53{]} & 0.152 & 2.32 {[}2.23,2.41{]} & 3.15 {[}3.06,3.24{]} & $\leq$0.001 & 2.80 {[}2.52,3.05{]} & 3.14 {[}2.84,3.48{]} & 0.176 \\
1990 & 3.39 {[}3.30,3.47{]} & 2.93 {[}2.85,3.02{]} & $\leq$0.001 & 3.08 {[}2.98,3.19{]} & 3.78 {[}3.68,3.88{]} & $\leq$0.001 & 3.72 {[}3.36,4.05{]} & 3.76 {[}3.38,4.17{]} & 0.858 \\
2000 & 4.21 {[}4.10,4.31{]} & 3.29 {[}3.19,3.40{]} & $\leq$0.001 & 3.83 {[}3.72,3.93{]} & 4.24 {[}4.11,4.34{]} & $\leq$0.001 & 4.62 {[}4.16,5.02{]} & 4.22 {[}3.75,4.68{]} & 0.349 \\
2010 & 5.01 {[}4.79,5.23{]} & 3.54 {[}3.38,3.73{]} & $\leq$0.001 & 4.55 {[}4.38,4.74{]} & 4.56 {[}4.34,4.75{]} & 0.978 & 5.49 {[}4.90,6.03{]} & 4.55 {[}4.03,5.06{]} & 0.049 \\
2016 & 5.52 {[}5.19,5.88{]} & 3.67 {[}3.44,3.92{]} & $\leq$0.001 & 5.00 {[}4.77,5.31{]} & 4.72 {[}4.43,4.99{]} & 0.196 & 6.05 {[}5.36,6.72{]} & 4.70 {[}4.13,5.29{]} & 0.010 \\
\bottomrule
\end{tabular}
\end{table}
\end{landscape}
}

\hypertarget{sensitivity-analyses-1}{%
\subsubsection{Sensitivity analyses}\label{sensitivity-analyses-1}}

The estimate of the DIC for each sensitivity analysis, by sex, is shown in
\textbf{Table \ref{tab:sensitivity-analysis-DIC}} (Supplementary Material 2)
along with the estimate for the selected model. The models in each sensitivity
analysis were less compatible with the data (greater value of DIC) compared to
the selected model, except for the marginal improvement in DIC when female
ex-smokers who had quit after the age of 40 years were allowed to switch to
reporting as never.

The proportion that initiated smoking did not appear sensitive to the source of
prior hazard ratios (being the CPS-I data), the use of a nonlinear cohort effect
in the quit rate (as opposed to calendar year), nor allowing those who quit
after age 40 years to report as never. There were small increases in the
proportion in the sensitivity analyses where: no ex-smokers were allowed to
switch to reporting as never; the survey weights were ignored, and; the
age at completed initiation was raised to 25 years. The proportion that initiated by age
20 years amongst those born in
1996 for all but the analysis of
sensitivity to age at complete initiation is shown in \textbf{Table
\ref{tab:sensitivity-analysis-proportion-initiate}} (Supplementary Material
2). For the analysis with an age at completed initiation of
25 years, the proportion of those
born in the year 1991 that initiated is shown in \textbf{Table
\ref{tab:first-age-analysis-proportion-initiate}} (Supplementary Material 2)
along with the sample from the selected model.

The sample of the quit rate in 2016 given the varying the assumptions tested in
sensitivity analysis is summarised in \textbf{Table
\ref{tab:sensitivity-analysis-quit-rate}} in the (Supplementary Material 2) by
sex and at age 30, 50 and 70 years. The greatest sensitivity was to the use of a
non-linear cohort effect in the model of the quit rate compared to a non-linear
calendar year effect as in the main analysis, resulting in a greater quit rate
across all ages and for both men and women, however this was less compatible
with the survey data than the main analysis according to the DIC. For all the
other variations in the models tested, the effects were mostly seen at older
ages, with one exception. The estimated quit rate at younger ages decreased
marginally if no ex-smokers were allowed to switch to reporting as never. At
older ages, the quit rate increased for both men and women when the prior hazard
ratios of mortality for smokers and ex-smokers was taken from the CPS-I study.
Amongst women, but not men, at older ages, the estimated quit rate increased
marginally if assumptions about ex-smokers being able to report as never smokers
were varied.

The sample of the rate of switching to reporting as a never smoker was steady
across the sensitivity analyses. The samples are summarised in \textbf{Table
\ref{tab:sensitivity-analysis-reversion-rate}} (Supplementary Material 2) by
sex and age-at-quit group. In the analysis allowing those who quit after age 40,
only female ex-smokers in this age-at-quit category appeared to switch at a rate
similar to age-at-quit 30-39 years.

The posterior of the age-standardised hazard ratio of mortality for smokers and
ex-smokers in each sensitivity analysis (except for the analysis using the
CPS-I data), along with the selected model from the main analysis, is provided
in \textbf{Table \ref{tab:hazard-ratio-in-sensitivity-analysis}} (Supplementary
Material 2), along with the overlap statistic between the prior and the
posterior. The posterior hazard ratio was consistent with the main analysis
except for a marginal increase in the ex-smoker hazard ratio when ex-smokers
were not allowed to switch to reporting as never.

\hypertarget{discussion}{%
\section{Discussion}\label{discussion}}

\hypertarget{bayesian-calibration-of-simulation-models}{%
\subsection{Bayesian calibration of simulation models}\label{bayesian-calibration-of-simulation-models}}

We have presented an outline of the Bayesian calibration of simulation models
common to several sciences. A Bayesian calibration is enabled by:
identifiability analyses; quantifying prior information about parameters;
statistical modelling of the observation process that was used to collect data;
Markov Chain Monte Carlo methods; model selection guided by information
criteria and other measures of predictive (or otherwise) performance;
validation; and, lastly, effective communication of the predictions and the
uncertainty that can be accounted for in the Bayesian framework.

Two of our steps focus on identifiability analyses and, within the Bayesian
framework, these can inform data collection and evidence gathering, can inform
the selection of the prior, and can signal a high risk of non-convergence of the
procedures used in later steps. The purpose of identifiability analysis is no
longer confined to demonstrating a desirable trait, it also guides decisions in
calibration and can eliminate waste of computational effort, and often even a
non-identifiable model can be calibrated when a suitable prior is available
(what makes a prior suitable depends on the context and philosophical attributes
listed by Gelman and Hennig, 2017 \citep{Gelman2017}).

Expertise across a number of disciplines will be required to succeed in
formulating and parameterising both the simulation model and the model of the
observation process used to collect data, and the vital step of constructing a
map between the two. The expertise may be spread across mathematicians, computer
scientists, and statisticians, as well as the need for experts or stakeholders
in the process being modelled.

While we embedded some Bayesian framework-specific consideration of internal
validity in the process, e.g.~measures of predictive performance, we did not
explicitly include \emph{face} or \emph{external validity}. Menzies et al. \citep{Menzies2017}
stated ``\ldots{} policy choice will not wait on all potential issues to be resolved.
In this context, calibration should be considered an exercise in creating a
reasonable model that produces valid evidence for policy'', which suggests that
while not all issues need be resolved, some demonstration of validity belongs in
calibration. Model validation is an evolving practise requiring ongoing research
\citep{Dahabreh2017}, and formal Bayesian frameworks have been developed
\citep{Bayarri2007, Collis2017}.

Estimates and predictions given by the sample of the posterior do not reflect
all sources of uncertainty. The uncertainty due to the calibration data being a
finite (random) sample is captured, while the uncertainty due to differences
between the real process and the model is not. Conveying the meaning of summary
statistics of the sample to the audience does not require that they are reported
as ``estimates of (a statistic) of (some) posterior distribution''. Using a cell
division rate as an example, some audiences would better understand the
statement ``90\% of our samples of the cell division rate were inside the interval
{[}0.02,0.05{]} events per hour'' than the statement ``We estimated that the 90\%
posterior credible interval of the cell division rate was {[}0.02, 0.05{]}''.
Discussion of which sources of uncertainty are accounted for in these
statements will be required regardless of which statement is used, however the
latter statement describes features that only some audiences may recognise or
care about.

\hypertarget{estimates-of-australian-smoking-behaviour}{%
\subsection{Estimates of Australian smoking behaviour}\label{estimates-of-australian-smoking-behaviour}}

We calibrated a model of smoking behaviour in Australia using the steps
outlined in \S \protect\hyperlink{bayesian-calibration-of-simulation-models-tutorial}{Bayesian calibration of simulation models: tutorial}. The calibration process
synthesised multiple data sources including 26 smoking surveys, population
mortality data \citep{HMD2019}, and the 45 and Up cohort study \citep{45AndUp2007}, and we
obtained samples from the model of: the predicted prevalence for each survey;
the proportion of each birth cohort that initiated smoking; the rate of
quitting; and the rate that those who had quit long-term switched to reporting
as having never smoked. The median of these samples and the intervals given by
their 5\textsuperscript{th} and 95\textsuperscript{th} percentiles describe the range of values that
are most compatible with the data sources we used and they indicate the range
that we would expect in the population.

Our model allowed those that had quit smoking to report as having never smoked,
which is one of the mechanisms that may explain increases in the never-smoker
proportion observed at early ages within a cohort \citep{Vaneckova2021}. This
transition was more frequent for those who quit before age 30 years than later,
and we speculated that it would also be more common for those who had smoked
occasionally or at a lower intensity. The inclusion of this pathway decreased
the discrepancy in the proportion of those who quit at younger ages, and was
more compatible with the data than a model without this pathway. Removing this
pathway decreased the estimated proportion that initiated and increased the quit
rate at middle or older ages. Other models may obtain better estimates of the
initiation rates and quit rates by including this effect.

We modelled the discrepancy in a similar, but not identical, fashion to that of
Kennedy and O'Hagan, 2001 \citep{Kennedy2001}, and we estimated it with a widely
available and efficient procedure \citep{Wood2017}. We described bias and unexplained
variance by age and cohort for a more diverse set of outcomes (expected
proportions of; smoker vs non-smoker, ex-smoker amongst non-smokers, and those
who quit before age 30 years amongst ex-smokers) than was presented for the
United States CISNET Smoking History Generator \citep{Holford2014A} and the
Australian model upon which this work was based \citep{Gartner2009}. This report may
be a useful blueprint for assessing model performance in a manner that mutes the
effects of stochastic (or aleatory) uncertainty.

The estimated discrepancy in the proportion of ex-smokers who quit before age 30
years suggested the model was biased towards earlier ages at quitting. This
could be caused by an over-estimated quit rate, or an insufficient model of the
transition from ex-smoker to never smoker. We are not aware of any assessment of
this in other smoking behaviour models, and therefore it is unclear if this bias
is considered a significant problem. If this bias is indeed a problem then one
could, rather than revising the model and calibrating again, incorporate the
estimate of the discrepancy into predictions \citep{Bayarri2007}.

We observed that only 52\% of the
observed cross-sectional proportions (of never smokers, smokers and ex-smokers)
were contained in the 90\% equal-tailed intervals. This deficiency could be
caused by either an incorrect model of the survey data, for example our
ambivalence to between-survey effects, or that the model is lacking some detail
or effect. Changes of delivery and the sequence or questions included can affect
the response rate, missing response rate and the classification of a respondent,
and generate between-survey effects. Extensive changes were made to the
questionnaire in the NDSHS in 1998, and estimates of the prevalence of current
smokers and ex-smokers in the NDSHS and NCADASIS prior to 1998 seem like
outliers relative to the 1998 (and later) NDSHS and the CCV surveys conducted
contemporaneously, suggestive of between-survey effects which we did not account
for.

We briefly consider issues regarding \emph{face validity}. We did not account for
the effect of migration in and out of the population. Differences in smoking
prevalence between immigrants and emigrants, combined with enough migration,
would bias the estimates of the rates and the model predictions \citep{Gartner2009}.
Earlier cohorts of women initiated smoking later \citep{Vaneckova2021} and this
could cause under-estimation of the quit rate at younger ages in earlier
cohorts; although this biased appeared small in sensitivity analysis.

This study had some structural limitations. 1) We did not consider interactions
between calendar year and age in cessation rates, meaning we could not examine
differences in the long-term effects of social norms and tobacco policy between
age groups. 2) Those who quit at younger ages have a lower risk of
smoking-related mortality than those who quit at later ages, however our model
did not vary the hazard ratio of mortality by age-at-quit, therefore our
estimate of the quit rate at younger ages was likely biased. 3) Importantly, we
have not considered how smoking intensity varies across age and between cohorts;
a vital measure of exposure for use in analyses of health outcomes in the
population.

\hypertarget{conclusion}{%
\section{Conclusion}\label{conclusion}}

Bayesian calibration enables the use of multiple data sources to inform model
predictions, which is vital when a single data source either lacks critical
information or is too imprecise on its own. The calibration process is supported
by identifiability analysis to minimise waste of computational effort, and
the robustness of the results can be strengthened by model selection and
analysis of model discrepancy.

Our Australian smoking behaviour model was the first cohort-specific model of
the life-course of smoking that was calibrated using Bayesian statistics and
was the first to demonstrate the significant impact of recanting bias on the
rates of initiation and quitting daily smoking. In recent years, fewer
Australians within a birth cohort started daily smoking than at any time over
the study period, and the rate that people quit daily-smoking is at its highest.
There are signs of emerging differences between men and women, with fewer women
starting smoking than men, and women who smoke are now quitting at a higher rate
than men. This model can be used to forecast future smoking prevalence rates in
the Australian population to assess the potential impact of new or ongoing
interventions in tobacco control.

\hypertarget{acknowledgements}{%
\section*{Acknowledgements}\label{acknowledgements}}
\addcontentsline{toc}{section}{Acknowledgements}

This research was completed using data collected through the 45 and Up Study
(\href{https://www.saxinstitute.org.au}{www.saxinstitute.org.au}). The 45 and Up Study is managed by the Sax
Institute in collaboration with major partner Cancer Council NSW; and partners:
the Heart Foundation; NSW Ministry of Health; NSW Department of Communities and
Justice; and Australian Red Cross Lifeblood. We thank the many thousands of
people participating in the 45 and Up Study.

Data linkage of 45 and Up Study and NSW Registryof Births Deaths and Marriages
performed by the NSW Ministry of Health's Centre for Health Record Linkage
(CHeReL; \href{https://www.cherel.org.au}{www.cherel.org.au}).

The authors acknowledge the Australian Data Archive for providing the following
datasets, and declare that those who carried out the original analysis and
collection of the data bear no responsibility for the further analysis or
interpretation of them:

\begin{itemize}
\tightlist
\item
  National Drug Strategy Household Survey 1995, 1998, 2001, 2004, 2007, 2010,
  2013, and 2016.
\item
  Victorian Drug Strategy Household Survey 1995.
\item
  National Campaign Against Drug Abuse and Social Issues Survey 1991, and
  1993.
\item
  Victorian Drug Household Survey (VDHS) 1993.
\item
  Social Issues Australia Survey 1985.
\item
  Risk Factor Prevalence Study (RFPS) 1980, 1983, and 1989.
\item
  Cancer Council Victoria Australian adult smoking surveys 1974, 1980, and
  1983.
\item
  Australian Gallup Polls (AGP) no. 158, 160, 168, and 193 (1962-1967).
\end{itemize}

The authors acknowledge Cancer Council Victoria for providing the following
data sets:

\begin{itemize}
\tightlist
\item
  Cancer Council Victoria Australian adult smoking survey data 1976, 1986,
  1989, 1992, and 1995.
\end{itemize}

Some of data analysis for this paper was generated using SAS software, Version
9.4 Release M6 of the SAS System for Windows (x64). Copyright \copyright 2018
SAS Institute Inc.~SAS and all other SAS Institute Inc.~product or service names
are registered trademarks or trademarks of SAS Institute Inc., Cary, NC, USA.

\hypertarget{funding}{%
\section*{Funding}\label{funding}}
\addcontentsline{toc}{section}{Funding}

This work was developed as part of an independent programme of work examining
the health impacts of e-cigarettes, funded by the Australian Government
Department of Health. EB is supported by a Principal Research Fellowship from
the National Health and Medical Research Council of Australia (reference:
1136128).

\hypertarget{bibliography}{%
\section*{Bibliography}\label{bibliography}}
\addcontentsline{toc}{section}{Bibliography}

\bibliography{CRDSHG-calibrate}

\newpage
\hypertarget{appendix-appendix}{%
\appendix}

\renewcommand{\thetable}{S1-\arabic{table}}
\renewcommand{\thefigure}{S1-\arabic{figure}}
\setcounter{table}{0}
\setcounter{figure}{0}

\hypertarget{supplementary-material-1-bayesian-simulation-model-calibration}{%
\section{Supplementary material 1: Bayesian simulation model calibration}\label{supplementary-material-1-bayesian-simulation-model-calibration}}

\hypertarget{metropolis-sampler}{%
\subsection{Metropolis Sampler}\label{metropolis-sampler}}

The Metropolis sampler produces a Markov chain whose stationary distribution is
a target distribution, \(f_\Theta\), given: a function \(y(\theta)\) which is
proportional to the target distribution (i.e.~\(y(\theta) \propto f_\Theta(\theta)\)), and; a symmetric `jump distribution' \(g_{\Theta'|\Theta}\).

To generate a chain of length \(N\):

\begin{itemize}
\tightlist
\item
  Start at an arbitrary \(\theta_0\).
\item
  For each \(t = 1, \ldots, N - 1\) perform the following:

  \begin{itemize}
  \tightlist
  \item
    Generate a candidate \(\theta'\) from the jump distribution
    \(g_{\Theta'|\Theta}( \theta' | \theta_{t-1} )\).
  \item
    Calculate \(\alpha\), the acceptance threshold (or ratio), given by
    \(y(\theta') / y(\theta_{t-1})\).
  \item
    Generate a number \(u\) from the distribution \(\operatorname{U}(0,1)\) and;
    if \(u < \alpha\) then `accept' the candidate and let \(\theta_t = \theta'\); otherwise `reject' the candidate and let \(\theta_t = \theta_{t-1}\).
  \end{itemize}
\end{itemize}

In a Bayesian calibration the target distribution is the posterior distribution
\(f_{\Theta|X}\), where \(\Theta\) is the vector of model parameters and \(X\) is the
calibration data. Provided one can compute the likelihood,
\(\mathcal{L}_{X|\Theta}\), and the prior, \(f_\Theta\), the sampler can be used to
construct chains with the posterior as the target distribution using the
substitution \(y(\theta) = \mathcal{L}_{X|\Theta} f_\Theta\). By Bayes' rule we
know that this function \(y(\theta) \propto f_{\Theta | X}\), i.e.~it is
proportional to the target distribution as required.

The algorithm will randomly walk through the sample space. The early iterations
from the finite-length chain are often discarded, as the early samples may be
biased by the choice of \(\theta_0\). The phase of the algorithm in which samples
are discarded is referred to as `burn in'. Further details on this and other
Markov Chain Monte Carlo (MCMC) samplers can be found in references such as
Brooks et al, 2011 \citep{Brooks2011}, and for Bayesian statistics in particular
Lunn et al, 2012 \citep{Lunn2012}, and Gelman et al, 2013 \citep{Gelman2013}.

\hypertarget{numerical-methods-for-practical-identifiability}{%
\subsection{Numerical methods for practical identifiability}\label{numerical-methods-for-practical-identifiability}}

\hypertarget{numerical-evaluation-of-profile-posterior}{%
\subsubsection{Numerical evaluation of profile posterior}\label{numerical-evaluation-of-profile-posterior}}

The profile posterior for the i\textsuperscript{th} component of a parameter \(\theta\) is the
function that maps each value \(t \in \mathbb{R}\) to the maximum value of the
posterior when \(\theta_i = t\), i.e.

\[ \begin{aligned}
    \operatorname{PP}_i (t|x) %
    &= \max_{\theta_{\setminus i}}
           \mathcal{L}_{ X | \Theta_{\setminus i},\Theta_i } \left(
               x | \theta_{\setminus i}, t
           \right)
           f_{ \Theta_{\setminus i},\Theta_i }( \theta_{\setminus i}, t ), \\
    &= \max_{\theta_{\setminus i}}
        f_{ \Theta_{\setminus i},\Theta_i | X }( \theta_{\setminus i}, t | x ),
\end{aligned} \]

where the subscript \(\setminus i\) indicates a vector with its i\textsuperscript{th} component
omitted, see equation 2.3 in Raue et al, 2013 \citep{Raue2013}.

An estimate \(\hat\theta_i\) is locally practically identifiable if the
posterior-based \(1 - \alpha\) `confidence' region is finite in extent. This can
be determined by considering three cases for the function \(\operatorname{PD}_i (t) = 2\operatorname{PP}_i(\hat\theta_i|x) - 2 \operatorname{PP}_i(t|x)\), with a
threshold \(\Delta_\alpha\), which is the \(1 - \alpha\) quantile of the \(\chi^2\)
distribution with degrees of freedom equal to the total number of parameters,
see figure 1 in Raue et al, 2013 \citep{Raue2013};

\begin{enumerate}
\def\labelenumi{\arabic{enumi}.}
\tightlist
\item
  Practically (locally) identifiable: A single trough occurs at
  \(\operatorname{PD}_i = 0\) and \(t = \hat\theta_i\), and the function increases
  to lie entirely above \(\Delta_a\) in both directions away from the trough.
\item
  Practically (locally) non-identifiable: A single trough occurs at
  \(\operatorname{PD}_i = 0\)
  and \(t = \hat\theta_i\), and the function increases to lie entirely above
  \(\Delta_a\) in at most one direction away from the trough.
\item
  Structurally (locally) non-identifiable: The function \(\operatorname{PD}_i\)
  is flat.
\end{enumerate}

The values of \(\operatorname{PD}_i\) may be obtained at evenly spaced values,
\(t_j\) (dropping subscript \(i\)), using a predictor-corrector approach. Let
\(\tilde\theta_j\) be the value of \(\theta\) which maximises the posterior while
the i\textsuperscript{th} element is fixed at \(t_j\). Starting from \(\tilde\theta_0 = \theta_{\text{MAP}}\) and moving away, a prediction \(\Delta \tilde\theta_{\text{pred}}\) for \(\Delta \tilde\theta_{j+1} = \tilde\theta_{j+1} - \tilde\theta_j\) can be found by solving for the local
optimum of the Taylor series approximation to the posterior at \(\tilde\theta_j\)
given by;

\[
    \operatorname{PP}_i(t_{j+1} | x) \approx %
        \operatorname{PP}_i(t_j | x) +
            \nabla_{\Theta} \Delta \tilde\theta_{\text{pred}} +
            {\Delta \tilde\theta_{\text{pred}}}^{\top}
                \operatorname{H} \left(
                    f_{\Theta | X} (\tilde\theta_j | x)
                \right)
                \Delta \tilde\theta_{\text{pred}},
\]

where \(\operatorname{H}\) is the Hessian matrix operator. The value of
\(\operatorname{H}\) may be approximated via Richardson extrapolation for \(j=0\)
(using the \texttt{hessian} function from the numDeriv package in R \citep{numDeriv2019})
and then updated prior to each predictor step for \(j \geq 1\) using the symmetric
rank-one (SR1) update formula. The local optimum is found by fixing \(\Delta \tilde\theta_{\text{pred},i} = t_{j+1}\) and equating zero to the gradient of
the right-hand side (with respect to \(\Delta \tilde\theta_{\text{pred}, \setminus i}\)). The Broyden, Fletcher, Goldfarb and Shanno optimisation
algorithm can be used as a corrector step to solve \(\Delta \tilde\theta_j\)
exactly \citep{Nocedal2006}. Using a predictor-corrector approach with SR1 update
may reduce computational effort significantly compared to no prediction step
(\(\Delta \tilde\theta_{\text{pred},\setminus i} = 0\)).

\hypertarget{overlap-statistic}{%
\subsubsection{Overlap statistic}\label{overlap-statistic}}

Another approach to assessing identifiability is to estimate an overlap
statistic for each parameter \citep{Garrett2000, Rutter2009}. This statistic can be
calculated for proper priors, but not improper priors. The overlap statistic
for a parameter \(\Theta\) is:

\[
    T_{\text{overlap}} = %
        \int_\Theta %
            \operatorname{min} \left( %
                f_\Theta(\theta), f_{\Theta | X}(\theta | x) %
            \right) \, %
        \operatorname{d} \theta.
\]

When a sufficient sample of the posterior has been obtained, this statistic
can be estimated by applying the mid-point algorithm to the kernel density
estimate of the posterior. The heuristic value of 0.35 or less can be used as a
reference for evidence of `weak' practical identifiability \citep{Garrett2000}.

\newpage
\renewcommand{\thetable}{S2-\arabic{table}}
\renewcommand{\thefigure}{S2-\arabic{figure}}
\setcounter{table}{0}
\setcounter{figure}{0}

\hypertarget{supplementary-material-2-estimates-of-australian-smoking-behaviour}{%
\section{Supplementary material 2: Estimates of Australian smoking behaviour}\label{supplementary-material-2-estimates-of-australian-smoking-behaviour}}

\hypertarget{formulation}{%
\subsection{Formulation}\label{formulation}}

\hypertarget{spline-terms}{%
\subsubsection{Spline terms}\label{spline-terms}}

We denote the initial (at age 20 years) proportion of
ex-smokers, amongst current and ex-smokers, as \(P_F\) and the proportion of a
cohort that initiated as \(P_I\) (one minus the proportion of never smokers). We
modelled the logit-transformed proportion that initiated as a spline-function
of the birth year with parameter vector \(Z_I\) representing the coefficients of
the spline, i.e.~for birth year \(c\);

\[ \log \frac{ P_I(c) }{ 1 - P_I(c) } = Z_{I,0} + f_I(c; Z_I) \]

for the intercept term \(Z_{I,0}\) and \(f_I\) a natural cubic spline.

We modelled the log-transformed quit rate as the sum of age and calendar year
(period) terms. Denote the coefficients of the respective splines as vectors
\(Z_{Q,\text{age}}\) and \(Z_{Q,\text{year}}\), and these combined to form the
parameter vector \(Z_Q = \begin{bmatrix} Z_{Q,\text{age}} & Z_{Q,\text{year}} \end{bmatrix}\). For age \(a\) and calendar year \(p\) the quit rate \(\lambda_Q\)
satisfied;

\[
    \log \lambda_Q(a,p) = Z_{Q,0} +
        f_{Q,\text{age}}(a; Z_{Q,\text{age}}) +
            f_{Q,\text{year}}(p; Z_{Q,\text{year}})
\]

for the intercept term \(Z_{Q,0}\) and two natural cubic splines
\(f_{Q,\text{age}}\) and \(f_{Q,\text{year}}\).

The log-transformed values of the rate of transition to the reporting-as-never
state (\(\lambda_R\)) were denoted as the parameter vector \(Z_R\).

The model parameters were then the (row) vector \(Z = \begin{bmatrix} P_F & Z_{I,0} & Z_{Q,0} & Z_I & Z_Q & Z_R \end{bmatrix}\) along with the age-group
specific hazard ratios \(HR_C\) and \(HR_F\).

The interior knots were placed at equal intervals in the following ranges:

\begin{itemize}
\tightlist
\item
  From year 1910 to 1997 for the
  birth year-effect spline, \(f_I\), in the proportion that initiate.
\item
  From age 20 to 100 years for the
  age-effect spline, \(f_{Q,\text{age}}\), in the quit rate.
\item
  From years 1930 to 2017 for the
  calendar year-effect spline \(f_{Q,\text{year}}\), in the quit rate.
\end{itemize}

\hypertarget{model-specifications}{%
\subsubsection{Model specifications}\label{model-specifications}}

We defined eight models with different numbers
of knots and with or without the effect that ex-smokers could report as a never
smoker, summarised in \textbf{Table \ref{tab:model-specification-summary}}. These
models are nested in the sense that the number of parameters describing an
effect always increases, but because the knot locations may vary between models
they are not nested in the sense that the models lie within subspaces of one
another.

\begin{table}
\caption{\label{tab:model-specification-summary}Specification of each candidate model in the Bayesian calibration of the Australian smoking behaviour model. Shown are the numbers of equi-spaced internal knots in each of the natural cubic splines; the birth year-effect in the proportion that initiate \(f_I\), the age-effect in the quit rate \(f_{Q,\text{age}}\), and the calendar year-effect in the quit rate \(f_{Q,\text{year}}\), along with whether ex-smokers who quit before age 40 years are able to report as a never smoker.}
\begin{tabular}{
  >{\raggedright\arraybackslash\bfseries}p{(\columnwidth - 12\tabcolsep) * \real{0.09}}
  >{\raggedright\arraybackslash}p{(\columnwidth - 12\tabcolsep) * \real{0.52}}
  >{\raggedleft\arraybackslash}p{(\columnwidth - 12\tabcolsep) * \real{0.03}}
  >{\raggedleft\arraybackslash}p{(\columnwidth - 12\tabcolsep) * \real{0.10}}
  >{\raggedleft\arraybackslash}p{(\columnwidth - 12\tabcolsep) * \real{0.10}}
  >{\raggedleft\arraybackslash}p{(\columnwidth - 12\tabcolsep) * \real{0.15}}}
\toprule
Model & \bfseries Description & \(f_I\) & \(f_{Q,\text{age}}\) &
\(f_{Q,\text{year}}\) & \bfseries Allow report-as-never \\
\midrule
Null & Constant proportion that initiate, constant quit rate, and no switching to never smoker for those who quit. & 0 & 0 & 0 & No \\
A & As `Null' +2 d.f. to birth year spline-effect in proportion that initiated. & 2 & 0 & 0 & No \\
B & As `A' +2 d.f. to age spline-effect in quit rate. & 2 & 2 & 0 & No \\
C & As `B' +1 d.f. to calendar year spline-effect in quit rate. & 2 & 2 & 1 & No \\
D & As `C' and allowed those who quit before age 40 years to switch to never smoker. & 2 & 2 & 1 & Yes \\
E & As `D' +1 d.f. to calendar year spline-effect in quit rate. & 2 & 2 & 2 & Yes \\
F & As `E' +1 d.f. to birth year spline-effect in proportion that initiated. & 3 & 2 & 2 & Yes \\
G & As `F' +1 d.f. to birth year spline-effect in proportion that initiated. & 4 & 2 & 2 & Yes \\
\bottomrule
\end{tabular}
\end{table}

\hypertarget{structural-identifiability-of-the-smoking-behaviour-model}{%
\subsubsection{Structural identifiability of the smoking behaviour model}\label{structural-identifiability-of-the-smoking-behaviour-model}}

To select identifiable parameters to calibrate we informally examined
structural identifiability. We considered the following simplified model for one
cohort where;

\begin{itemize}
\tightlist
\item
  recent quitters were indistinguishable from current smokers;
\item
  reporting-as-never smokers were indistinguishable from never smokers;
\item
  there were no `age-at-quit' categories; and
\item
  never-smoker mortality was zero via an integrating factor.
\end{itemize}

The first two simplifications reflect the survey data, and we assert that the
second two did not materially impact our informal examination. The following
equations describe the age-evolution of the expected proportions in the cohort:

\[ \begin{aligned}
    \frac{\mathrm{d} N}{\mathrm{d} a} &= \lambda_{R} F, \\
    \frac{\mathrm{d} C}{\mathrm{d} a} &= %
        -\left( \lambda_Q + \mu_C \right) C, \text{ and } \\
    \frac{ \mathrm{d} F }{ \mathrm{d} a } &= %
        \lambda_Q C - \left( \lambda_{R} + \mu_F \right) F,
\end{aligned} \]

where \(N\), \(C\), and \(F\) are the expected proportions of never, current and
ex-smokers. We assumed that the ideal outputs are differentiable quantities \(N\),
\(C\) and \(F\) for all values of the continuous age, \(a\). Our informal analysis
of identifiability followed: The first equation illustrates that growth in \(N\)
and the value of \(F\) identified the rate of reversion. The remaining two
equations contained three unknowns \(\lambda_Q\), \(\mu_C\) and \(\mu_F\), and any
values of these denoted \(\tilde \lambda_Q\), \(\tilde \mu_C\) and \(\tilde \mu_F\),
respectively, could be replaced with \(\tilde \lambda_Q + \epsilon\), \(\tilde \mu_C - \epsilon\) and \(\tilde \mu_F - \epsilon C / F\), for any function of age
\(\epsilon\), and the equations would still be satisfied. Therefore these three
parameters were an unidentifiable set. With additional data on at least one of
the excess mortality rates, the quit rate (and the other mortality rate) would
have been identifiable. The initial state of the cohort was identified by the
values of \(N\), \(C\) and \(F\) at the starting age of the simulation. The
identifiable parameters were therefore \(Z\) and, at most, one of \(HR_C\) or
\(HR_F\).

\hypertarget{detailed-prior-distribution}{%
\subsubsection{Detailed prior distribution}\label{detailed-prior-distribution}}

We assumed that the joint prior could be decomposed into mutually independent
priors for; the hazard ratios (jointly \(HR = \begin{bmatrix} HR_C & HR_F \end{bmatrix}\)), the proportion \(P_F\), each of the coefficient vectors \(Z_I\)
and \(Z_Q\), and each rate of switching to reporting-as-never in \(Z_R\).

We used a multivariate normal distribution as an informative prior for the
logarithm of the hazard ratios in \(HR\). The mean and covariance of the
distribution were provided by Cox regressions using the 45 and Up Study,
outlined in the main text. The analysis was performed with SAS software Version
9.4. The mean was provided by the maximum likelihood estimate (MLE) of the log
hazard ratio, and the covariance was estimated by the negative of the inverse of
the estimated observed information \citep{SAS2018}.

We assigned a normal distribution as a prior for \(\operatorname{logit} (2 P_F)\)
with mean \(\operatorname{logit} (0.3)\) and \(\mathbb P (\operatorname{logit} 2 P_F < 0.02) = 5\%\). We assigned each rate of switching to reporting as never,
\(\lambda_R\), the prior \(1 / \sqrt{\lambda_R}\), chosen for being the Jeffreys
prior for an exponential random variable, noting that it is \emph{not} a Jeffreys
prior given the survey data.

We used multivariate normal priors for the parameter vectors \(\begin{bmatrix} Z_{I,0} & Z_I \end{bmatrix}\) and \(\begin{bmatrix} Z_{Q,0} & Z_Q \end{bmatrix}\),
which provide a small penalty for values away from zero, but were intended to
provide little information compared to the survey data. The priors were of the
form \(\mathcal{N} (0, (X^\top W X)^{-1} w / \sigma^2)\), where \(X\) was a basis
matrix, \(W\) was a diagonal matrix of weights with entries given by the effective
sample size for the corresponding row in \(X\), \(w\) was the sum of the elements of
\(W\), and \(\sigma\) was a nuisance parameter. The value of \(X\) for the prior of
\(\begin{bmatrix} Z_{I,0} & Z_I \end{bmatrix}\) was given by the evaluating the
basis functions of the spline \(f_I\) at the birth year of a never smoker up to
age 25 years, and a column of ones for the intercept
term. The value of \(X\) for the prior of \(\begin{bmatrix} Z_{Q,0} & Z_Q \end{bmatrix}\) was similarly determined by evaluating the basis functions for
the sum of the age and calendar effects, \(f_{Q,\text{age}}\) and
\(f_{Q,\text{year}}\), of smokers in the surveys. We assigned each nuisance
parameter the prior \(1 / \sigma\).

\hypertarget{graphical-model-of-smoking-behaviour-calibration}{%
\subsubsection{Graphical model of smoking behaviour calibration}\label{graphical-model-of-smoking-behaviour-calibration}}

\begin{figure}
\centering
\includegraphics{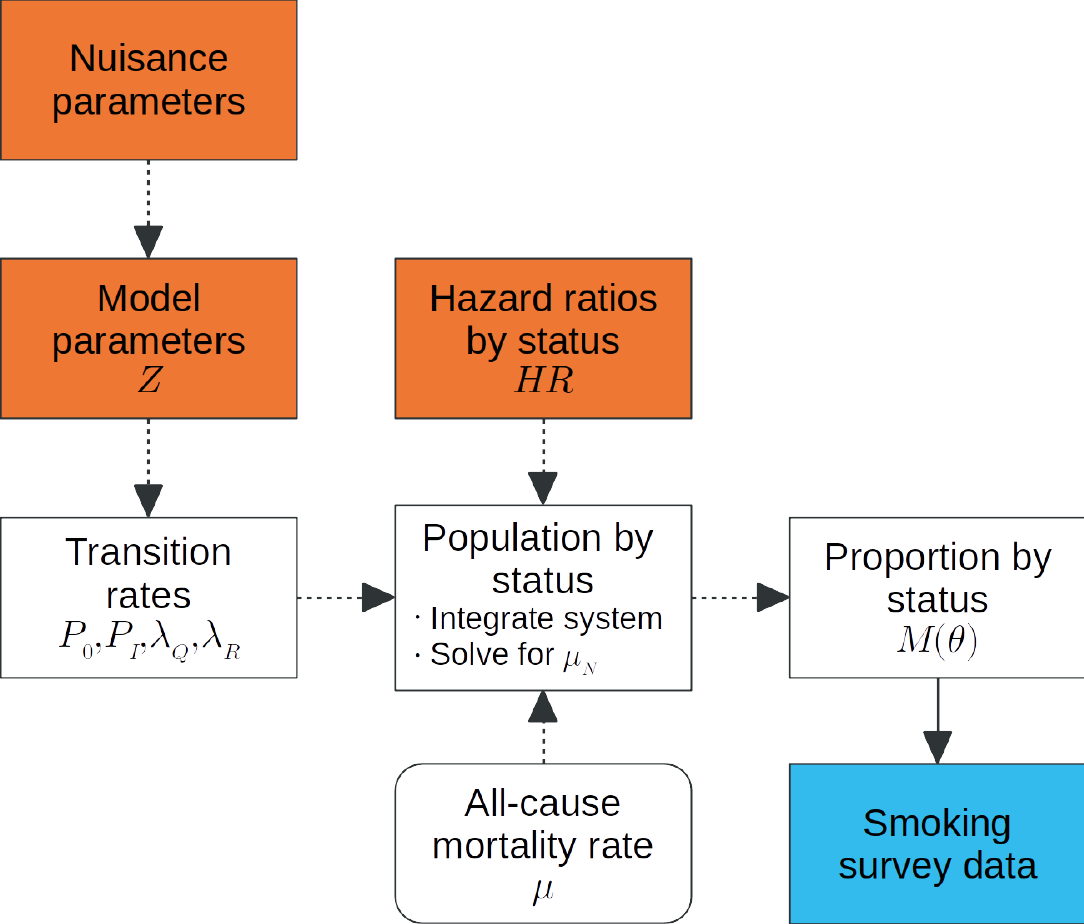}
\caption{\label{fig:calibration-graphical-model}Directed acyclic graph outlining the dependency structure in the calibration of Australian smoking behaviour model. Nodes represent quantities, arrows represent parent-child relationships, and quantities are conditionally independent of all other nodes given their parents and descendants. Model parameters \(Z\), hazard ratios \(HR\), and quantities \(P_0\), \(P_I\), \(\lambda_Q\), \(\lambda_R\), are defined in main text. All-cause mortality rate \(\mu\) treated as constant; \(\mu_N\) is the never-smoker mortality rate. The values given by \(M(\theta)\) correspond to the natural parameters of the likelihood of the smoking survey data.}
\end{figure}

We assumed the dependence structure for the calibration as presented in the
directed acyclic graph in \textbf{Figure \ref{fig:calibration-graphical-model}}.
Parent-child relationships are indicated by directed arrow between nodes in the
graph, nodes were conditionally independent of all other nodes except their
parents and descendants, and the value of a parents typically `generated' the
values of its children \citep{Jackson2015}. The nuisance parameters, hazard ratios,
and the all-cause mortality data had no parents. The former two were random
variables with priors, while the latter was treated as a constant in the model.
The model parameters that describe the initial state and the transition rate
between states were children of the nuisance parameters. In turn, the initial
state and transition rates themselves were children of the parameters, which
were governed by the functions discussed in \S \protect\hyperlink{parameterisation}{Parameterisation}. The age-specific population in each state (within a cohort)
and the never smoker mortality, solved simultaneously as described in \S
\protect\hyperlink{single-cohort-equations}{Single cohort equations} below, were children of; the
rates; the hazard ratios, and; the mortality data. The values of \(M(\theta)\),
which was the map between the model parameters and the likelihood of the survey
data, were given by scaling the expected population to an expected proportion
and was a child of the population node. Finally, the survey data was the only
child node of the proportions node and was distributed as per \S \protect\hyperlink{survey-data-likelihood}{Survey
data likelihood}.

\hypertarget{single-cohort-equations}{%
\subsubsection{Single cohort equations}\label{single-cohort-equations}}

For each smoking status category, the expected population size in a cohort was
found by step-wise numerical integration of the delay-differential
equations;

\[ \begin{aligned}
    \frac{\mathrm{d} N}{\mathrm{d} a} &= -\mu_N N \\
    \frac{\mathrm{d} C}{\mathrm{d} a} &= %
        -\left( \lambda_Q + \mu_C \right) C, \\
    \frac{ \mathrm{d} Q_j }{ \mathrm{d} a } &= %
        \mathbb{I}_{[a_j,a_{j+1})} \lambda_Q C -
            \left[ \mathbb{I}_{[a_j,a_{j+1})} \lambda_Q S_Q C \right]_{a-k} -
                \mu_C Q,
\end{aligned} \]

where \(S_Q(a) = \exp \left( -\int_{a}^{a+k} \mu_C (s) \, \mathrm{d} s \right)\)
was the survival of recent quitters at age \(a+k\) conditional on survival
to age \(a\);

\[ \begin{aligned}
    \frac{ \mathrm{d} F_j }{ \mathrm{d} a } &= \
        \left[ \mathbb{I}_{[a_j,a_{j+1})} \lambda_Q S_Q C \right]_{a-k} -
             \left( \lambda_{R,j} + \mu_F \right) F_j,
    \text{ and } \\
    \frac{ \mathrm{d} R_j }{ \mathrm{d} a } &= \lambda_{R,j} F_j - \mu_N R_j,
\end{aligned} \]

starting with an initial population at age \(\hat{a} =\) 20. The initial population was made up of never smokers,
current smokers, and recent quitters or ex-smokers belonging to the first
age-at-quit interval (shown in \textbf{Figure \ref{fig:structure-of-model}}).
The initial number of recent quitters was determined using the following
assumptions:

\begin{itemize}
\tightlist
\item
  The recent-quitter category was empty \(k=2\) years prior to the initial age,
  \(\hat{a}\);
\item
  No excess mortality occurred in smokers prior to the initial age;
\item
  The quit rate in the \(k=2\) years prior to the initial age was constant and
  equal to the quit rate at the initial age.
\end{itemize}

\hypertarget{integration-by-quadrature}{%
\paragraph{Integration by quadrature}\label{integration-by-quadrature}}

The system of equations, for a given \(\mu_N\), was solved by nested Gaussian
quadrature rules, using the scale \(\exp \left( -\int_{\hat a}^{a} \mu_N(s) \, \mathrm{d} s \right)\) to reduce the number of equations by one; the scaled
never-smoker population size was trivially a constant. The current smoker
population, \(C\), was solved by Gauss-Legendre quadrature at each integer age, as
the mortality rate was assumed constant within each year. Then, the
recent-quitter and ex-smoker populations, \(Q_j\) and \(F_j\), for each age-at-quit
category, could be determined at each integer age with Gauss-Legendre
quadrature, interpolating the solution of \(C\) as needed. Finally, the
reporting-as-never population, \(R_j\), was calculated in the same fashion,
interpolating \(F_j\) as needed.

The value of \(\mu_N\) was found by simulataneously solving for its relationship
to the population mortality \(\mu\) using the hazard ratios. This was achieved via
the secant method to a tolerance set close to the size of the error expected in
the quadrature rules.

\hypertarget{survey-data-likelihood}{%
\subsubsection{Survey data likelihood}\label{survey-data-likelihood}}

An expression for the likelihood function for the survey data would be trivial
to formulate were it not for the non-random sampling that occurred either by
survey design, e.g.~surveys that used multi-stage stratification or
quota-sampling designs, or through effects such as non-response. The bias due to
non-random sampling has been reduced in reported estimates by weighting the
responses. We incorporated the weights for each participant, where supplied, in
the following formulation of the likelihood:

\begin{itemize}
\item
  The proportions in each survey-category were vectors, denoted \(p_i\) with the
  index \(i\) to each cell in the cross-table of the data by sex, age, birth
  year and survey.
\item
  Each \(p_i\) was independent.
\item
  Each \(p_i\) was observed with the effective sample size \(n_{\text{eff},i} = \left( \sum_j w_j \right)^2 / \sum_j (w_j)^2\) where \(j\) was the index for
  each participant (within the i\textsuperscript{th} cell in the cross-table).
\item
  The likelihood for each \(p_i\) was:

  \[
      f_{P_i | \Theta}(p_i | \theta) = \operatorname{Dir}
          \left( M_i(\theta); 1 + p_i n_{\text{eff}, i} \right)
  \]

  where \(M_i(\theta)\) was the expected value of each proportion for the model
  by age, sex, and birth year, and \(\operatorname{Dir}\) denotes the Dirichlet
  distribution.
\end{itemize}

We found the expected proportions by numerically solving the governing
equations detailed in the preceding section.

The predicted counts of smoking status for a given sex, age and birth year that
was sampled \(n_i\) times in the survey (where \(i\) is the same index as above)
were modelled as Dirichlet-Multinomial random variables. We chose the parameters
of the distribution for each count such that the variance of the counts would
match that of a multinomial with the effective sample size of the cell, i.e.~for
the count \(x_i\);

\[
    f_{X_i | \Theta}(x_i | \theta) =
        \operatorname{DirMult}(x_i; n=n_i, \alpha=M_i(\theta) \alpha_{0,i} ),
\]

where \(\alpha_{0,i} = n_i (n_{\text{eff},i} - 1) / (n_i - n_{\text{eff},i})\).

\hypertarget{metropolis-within-gibbs-sampler}{%
\subsection{Metropolis-within-Gibbs sampler}\label{metropolis-within-gibbs-sampler}}

We used the Metropolis-within-Gibbs algorithm to sample from the joint posterior
of \(HR\), \(Z\), and the nuisance parameters. Each of these was the basis of a
block in the Gibbs sampler. We used the asymptotic distributions of \(HR\) at its
MLE, and \(Z\) and the nuisance parameters at their maximum a posteriori (MAP)
estimate, to sample a starting position for each chain, and as the initial
covariance for the (normal) jumping distribution for each block.
Five chains were
simulated, with the starting point sampled from the aforementioned distributions
with covariance multiplied by a factor of five. The first 1600 samples were used to determine a scale and
covariance for the jumping distribution, targeting an acceptance rate of 0.234,
and then discarded. We monitored convergence using the potential variance
reduction of the discarded (burn-in) and all samples achieved \(\hat{R} < 1.3\).
Sampling continued thereon until an estimated effective sample size of
50 was obtained \citep[see][\S11 equation 11.8]{Gelman2013}. We
culled each chain to 40 evenly-spaced samples,
for a total of 200 samples from the posterior, to reduce the
effort in downstream calculations.

The acceptance ratio for each block in the Gibbs sampler was formulated with the
aid of the dependence structure shown in \textbf{Figure
\ref{fig:calibration-graphical-model}}. The dependence structure simplified
the evaluation of the acceptance ratio for the jump in the nuisance parameters
as it could be evaluated independently of the likelihood of the survey data.

We outline below the algorithm we used to update the jump distribution for each
block at the end of the burn-in phase. We combined the covariance of the burn-in
sample with the initial covariance of each block's jump distribution to obtain
jump distributions that would sample a normal approximation of the posterior
more efficiently.

\hypertarget{modified-burn-in-phase-for-calibration}{%
\subsubsection{Modified burn-in phase for calibration}\label{modified-burn-in-phase-for-calibration}}

We modified the Metropolis-within-Gibbs sampler during the burn-in phase to
estimate a more efficient jump distribution for each block in the post-burn-in
phase. The burn-in phase jump distributions (shared across chains \(m =\)
5 chains) were fixed within each sub-interval, of width
\(\Delta_n\), in each chain (we used \(\Delta_n =\) 10). The block-specific jump
distributions in the k\textsuperscript{th} sub-interval were given by zero-centred multivariate
normal distributions with covariances equal to estimated block-specific
posterior covariances, \(\hat{\Sigma}_k\), multiplied by a block-specific factor,
\(\hat{\beta}_k\). Initial estimates of the posterior covariances were given by
\(\hat{S}_0 = \Sigma_{\text{MAP}}\), the asymptotic estimate given by the
negative inverse Hessian of \(f_{\Theta|X}\) evaluated at \(\theta_{\text{MAP}}\),
or, with a similar definition, \(\hat{S}_0 = \Sigma_{\text{MLE}}\) for the hazard
ratios.

The k\textsuperscript{th} estimates of the block-specific posterior covariances were drawn from
the inverse Wishart distribution \(W^{-1} (\hat{\Psi}_k, mk\Delta_n)\) where;
\(\hat{\Psi}_k = \hat{\Psi}_{k-1} + m\Delta_n \hat{S}_{k-1}\); \(\hat{\Psi}_0 = 0\);
and each \(\hat{S}_k\) was given by the pooled sample covariance from the
\(\Delta_n\) samples preceding the \(k\Delta_n +\) 1\textsuperscript{th} iteration across all
chains. The initial values of \(\beta_1\) for each block were given by
\(2.4/\sqrt{d}\) where \(d\) was the number of dimensions in the block. Each value
of \(\beta_k\) was given by randomly increasing or decreasing the corresponding
previous value, based on the probability that the (block-specific) acceptance
rate from pooling the \(\Delta_n\) iterations (preceding the \(k\Delta_n +\) 1\textsuperscript{th}
iteration) from each chain was greater than 0.234, using Pearson's chi-squared
statistic for a proportion.

Once the potential variance reduction measure over the last \(b\) sub-intervals
(we used \(b =\) 80) was found to be less than 1.3, a new
jump distribution was estimated which would be fixed throughout the post-burn-in
phase. The most recent \(b\Delta_n\) samples from each chain were pooled to
estimate the posterior covariance. The jump distribution's covariance was the
estimated posterior covariance multiplied by a point estimate, for an
acceptance rate of 0.234, from the regression of \(\beta_k\) onto the obtained
acceptance rates in the \(b\) preceding sub-intervals.

\begin{landscape}

\hypertarget{additional-calibration-results}{%
\subsection{Additional calibration results}\label{additional-calibration-results}}

\hypertarget{summary-of-profile-posterior-near-maximum-a-posteriori-estimates}{%
\subsubsection{Summary of profile posterior near maximum a posteriori estimates}\label{summary-of-profile-posterior-near-maximum-a-posteriori-estimates}}


\begin{table}[h]
\addtolength{\tabcolsep}{-4pt}
\centering
\caption{\label{tab:profile-posterior-confidence-summary}Highest confidence level for which the associated posterior-based confidence interval (see Raue et al, 2013 \citep{Raue2013}) of the \emph{maximum a posteriori} (MAP) estimate of a parameter was contained within the neighbourhood given by ± 7.1 standard deviations of the estimate's asymptotic distribution. Description of each parameter is provided in the main text. Description of each model (labelled Null, and A-G) provided in \textbf{Table \ref{tab:model-specification-summary}}. Highest level greater than 95\% indicates local practical identifiability.}
\begin{tabular}{
  >{\raggedright\arraybackslash\bfseries\small}p{(\columnwidth - 34\tabcolsep) * \real{0.060}}
  >{\raggedright\arraybackslash\small}p{(\columnwidth - 34\tabcolsep) * \real{0.050}}
  >{\raggedleft\arraybackslash\small}p{(\columnwidth - 34\tabcolsep) * \real{0.055}}
  >{\raggedleft\arraybackslash\small}p{(\columnwidth - 34\tabcolsep) * \real{0.055}}
  >{\raggedleft\arraybackslash\small}p{(\columnwidth - 34\tabcolsep) * \real{0.060}}
  >{\raggedleft\arraybackslash\small}p{(\columnwidth - 34\tabcolsep) * \real{0.045}}
  >{\raggedleft\arraybackslash\small}p{(\columnwidth - 34\tabcolsep) * \real{0.065}}
  >{\raggedleft\arraybackslash\small}p{(\columnwidth - 34\tabcolsep) * \real{0.065}}
  >{\raggedleft\arraybackslash\small}p{(\columnwidth - 34\tabcolsep) * \real{0.065}}
  >{\raggedleft\arraybackslash\small}p{(\columnwidth - 34\tabcolsep) * \real{0.065}}
  >{\raggedleft\arraybackslash\small}p{(\columnwidth - 34\tabcolsep) * \real{0.045}}
  >{\raggedleft\arraybackslash\small}p{(\columnwidth - 34\tabcolsep) * \real{0.060}}
  >{\raggedleft\arraybackslash\small}p{(\columnwidth - 34\tabcolsep) * \real{0.060}}
  >{\raggedleft\arraybackslash\small}p{(\columnwidth - 34\tabcolsep) * \real{0.065}}
  >{\raggedleft\arraybackslash\small}p{(\columnwidth - 34\tabcolsep) * \real{0.065}}
  >{\raggedleft\arraybackslash\small}p{(\columnwidth - 34\tabcolsep) * \real{0.060}}
  >{\raggedleft\arraybackslash\small}p{(\columnwidth - 34\tabcolsep) * \real{0.060}}}
\toprule
Sex &
\bfseries    Model &
\(\log \sigma_{\text{quit}}\) &
\(\log \sigma_{\text{init}}\) & 
\(\operatorname{logit} P_{\text{F}}\) &
\(Z_{\text{I},0}\) &
\(Z_{\text{I},\text{cohort},1}\) &
\(Z_{\text{I},\text{cohort},2}\) &
\(Z_{\text{I},\text{cohort},3}\) &
\(Z_{\text{I},\text{cohort},4}\) &
\(Z_{\text{Q},0}\) &
\(Z_{\text{Q},\text{age},1}\) &
\(Z_{\text{Q},\text{age},2}\) &
\(Z_{\text{Q},\text{year},1}\) &
\(Z_{\text{Q},\text{year},2}\) &
\(\log \lambda_{\text{R},1}\) &
\(\log \lambda_{\text{R},2}\) \\
\midrule
Women & Null & 84\% & 95\% & \textgreater95\% & \textgreater95\% & & & & & \textgreater95\% & & & & & & \\
      & A & 78\% & 95\% & \textgreater95\% & \textgreater95\% & \textgreater95\% & \textgreater95\% & & & \textgreater95\% & & & & & & \\
      & B & \textgreater95\% & \textgreater95\% & \textgreater95\% & \textgreater95\% & \textgreater95\% & \textgreater95\% & & & \textgreater95\% & \textgreater95\% & \textgreater95\% & & & & \\
      & C & 90\% & 84\% & \textgreater95\% & \textgreater95\% & \textgreater95\% & \textgreater95\% & & & \textgreater95\% & \textgreater95\% & \textgreater95\% & \textgreater95\% & & & \\
      & D & 85\% & 71\% & \textgreater95\% & \textgreater95\% & \textgreater95\% & \textgreater95\% & & & \textgreater95\% & \textgreater95\% & \textgreater95\% & \textgreater95\% & & \textgreater95\% & \textgreater95\% \\
      & E & 84\% & 65\% & \textgreater95\% & \textgreater95\% & \textgreater95\% & \textgreater95\% & & & \textgreater95\% & \textgreater95\% & \textgreater95\% & \textgreater95\% & \textgreater95\% & \textgreater95\% & \textgreater95\% \\
      & F & 90\% & 68\% & \textgreater95\% & \textgreater95\% & \textgreater95\% & \textgreater95\% & \textgreater95\% & & \textgreater95\% & \textgreater95\% & \textgreater95\% & \textgreater95\% & \textgreater95\% & \textgreater95\% & 92\% \\
      & G & 79\% & 68\% & \textgreater95\% & \textgreater95\% & \textgreater95\% & \textgreater95\% & \textgreater95\% & \textgreater95\% & \textgreater95\% & \textgreater95\% & \textgreater95\% & \textgreater95\% & \textgreater95\% & \textgreater95\% & 90\% \\
Men & Null & 84\% & 94\% & \textgreater95\% & \textgreater95\% & & & & & \textgreater95\% & & & & & & \\
    & A & 78\% & \textgreater95\% & \textgreater95\% & \textgreater95\% & \textgreater95\% & \textgreater95\% & & & \textgreater95\% & & & & & & \\
    & B & \textgreater95\% & \textgreater95\% & \textgreater95\% & \textgreater95\% & \textgreater95\% & \textgreater95\% & & & \textgreater95\% & \textgreater95\% & \textgreater95\% & & & & \\
    & C & 91\% & 85\% & \textgreater95\% & \textgreater95\% & \textgreater95\% & \textgreater95\% & & & \textgreater95\% & \textgreater95\% & \textgreater95\% & \textgreater95\% & & & \\
    & D & 85\% & 71\% & \textgreater95\% & \textgreater95\% & \textgreater95\% & \textgreater95\% & & & \textgreater95\% & \textgreater95\% & \textgreater95\% & \textgreater95\% & & \textgreater95\% & 63\% \\
    & E & 91\% & 77\% & \textgreater95\% & \textgreater95\% & \textgreater95\% & \textgreater95\% & & & \textgreater95\% & \textgreater95\% & \textgreater95\% & \textgreater95\% & \textgreater95\% & \textgreater95\% & 62\% \\
    & F & 87\% & 87\% & \textgreater95\% & \textgreater95\% & \textgreater95\% & \textgreater95\% & \textgreater95\% & & \textgreater95\% & \textgreater95\% & \textgreater95\% & \textgreater95\% & \textgreater95\% & \textgreater95\% & 34\% \\
    & G & 77\% & 71\% & \textgreater95\% & \textgreater95\% & \textgreater95\% & \textgreater95\% & \textgreater95\% & \textgreater95\% & \textgreater95\% & \textgreater95\% & \textgreater95\% & \textgreater95\% & \textgreater95\% & \textgreater95\% & 27\% \\
\bottomrule
\end{tabular}
\end{table}

\clearpage

\hypertarget{summary-of-prior-and-posterior-hazard-ratios}{%
\subsubsection{Summary of prior and posterior hazard ratios}\label{summary-of-prior-and-posterior-hazard-ratios}}

\begin{table}[h]
\centering
\caption{\label{tab:hazard-ratio-overlap}Hazard ratio of all-cause mortality in the Australian population by sex and 5-year age groups (45-49 years to 85-89 years) and the age group 90-99 years, comparing smokers and ex-smokers to never smokers. Shown are the median and the interval given by the 5\textsuperscript{th} and 95\textsuperscript{th} percentiles of; the asymptotic distribution of the maximum likelihood estimate of a Cox proportional hazard model applied to the 45 and Up Study cohort; and the posterior distribution given by the Australian smoking model from the main analysis. Also shown are the overlap statistics between the prior and posterior \citep{Garrett2000}.}
\begin{tabular}{
  >{\raggedright\arraybackslash\bfseries}p{(\columnwidth - 16\tabcolsep) * \real{0.11}}
  >{\raggedright\arraybackslash}p{(\columnwidth - 16\tabcolsep) * \real{0.11}}
  >{\raggedleft\arraybackslash}p{(\columnwidth - 16\tabcolsep) * \real{0.17}}
  >{\raggedleft\arraybackslash}p{(\columnwidth - 16\tabcolsep) * \real{0.14}}
  >{\raggedleft\arraybackslash}p{(\columnwidth - 16\tabcolsep) * \real{0.08}}
  >{\raggedleft\arraybackslash}p{(\columnwidth - 16\tabcolsep) * \real{0.17}}
  >{\raggedleft\arraybackslash}p{(\columnwidth - 16\tabcolsep) * \real{0.14}}
  >{\raggedleft\arraybackslash}p{(\columnwidth - 16\tabcolsep) * \real{0.08}}}
\toprule
 & &
    \multicolumn{3}{c}{\bfseries Women} &
    \multicolumn{3}{c}{\bfseries Men} \\
                  \cmidrule(lr){3-5} \cmidrule(l){6-8}
Status &
    \bfseries Age group &
    \bfseries 45 \& Up Study prior & \bfseries Posterior & \bfseries Overlap &
    \bfseries 45 \& Up Study prior & \bfseries Posterior & \bfseries Overlap \\
\cmidrule(r){1-2} \cmidrule(lr){3-5} \cmidrule(l){6-8}
Smoker & 45-49 & 3.37 {[}2.53,4.50{]} & 3.14 {[}2.46,4.27{]} & 0.86 & 4.01 {[}3.09,5.21{]} & 3.53 {[}2.73,4.56{]} & 0.69 \\
       & 50-54 & 3.20 {[}2.54,4.04{]} & 3.27 {[}2.45,4.10{]} & 0.91 & 4.07 {[}3.34,4.95{]} & 3.76 {[}3.05,4.56{]} & 0.74 \\
       & 55-59 & 3.49 {[}2.83,4.30{]} & 3.71 {[}3.07,4.56{]} & 0.81 & 4.02 {[}3.39,4.78{]} & 3.87 {[}3.29,4.80{]} & 0.86 \\
       & 60-64 & 3.61 {[}3.00,4.34{]} & 4.01 {[}3.31,4.72{]} & 0.60 & 3.64 {[}3.11,4.26{]} & 3.73 {[}3.20,4.32{]} & 0.90 \\
       & 65-69 & 3.98 {[}3.32,4.78{]} & 4.58 {[}3.89,5.42{]} & 0.50 & 3.14 {[}2.70,3.65{]} & 3.36 {[}2.92,3.94{]} & 0.70 \\
       & 70-74 & 3.31 {[}2.69,4.08{]} & 4.30 {[}3.55,5.13{]} & 0.33 & 3.18 {[}2.72,3.72{]} & 3.57 {[}3.04,4.20{]} & 0.53 \\
       & 75-79 & 2.12 {[}1.67,2.70{]} & 2.41 {[}1.93,3.19{]} & 0.66 & 2.42 {[}2.04,2.88{]} & 2.57 {[}2.15,3.06{]} & 0.78 \\
       & 80-84 & 2.44 {[}2.00,2.99{]} & 2.50 {[}2.03,2.94{]} & 0.92 & 1.96 {[}1.66,2.32{]} & 2.01 {[}1.71,2.33{]} & 0.92 \\
       & 85-89 & 1.30 {[}0.97,1.76{]} & 1.31 {[}0.97,1.77{]} & 0.96 & 1.55 {[}1.12,2.13{]} & 1.28 {[}0.92,1.66{]} & 0.61 \\
       & 90-99 & 0.98 {[}0.50,1.94{]} & 0.89 {[}0.57,1.45{]} & 0.83 & 1.54 {[}0.73,3.24{]} & 1.23 {[}0.83,1.96{]} & 0.70 \\
Ex-smoker & 45-49 & 1.10 {[}0.79,1.53{]} & 1.11 {[}0.79,1.55{]} & 0.94 & 1.28 {[}0.94,1.75{]} & 1.32 {[}0.91,1.75{]} & 0.92 \\
          & 50-54 & 1.44 {[}1.15,1.80{]} & 1.43 {[}1.12,1.79{]} & 0.94 & 1.16 {[}0.93,1.45{]} & 1.13 {[}0.91,1.41{]} & 0.92 \\
          & 55-59 & 1.57 {[}1.29,1.91{]} & 1.58 {[}1.30,1.93{]} & 0.96 & 1.35 {[}1.14,1.60{]} & 1.30 {[}1.14,1.48{]} & 0.84 \\
          & 60-64 & 1.44 {[}1.22,1.70{]} & 1.44 {[}1.22,1.70{]} & 0.95 & 1.36 {[}1.19,1.56{]} & 1.37 {[}1.17,1.56{]} & 0.94 \\
          & 65-69 & 1.61 {[}1.40,1.87{]} & 1.64 {[}1.40,1.86{]} & 0.94 & 1.50 {[}1.33,1.68{]} & 1.49 {[}1.30,1.66{]} & 0.93 \\
          & 70-74 & 1.32 {[}1.15,1.53{]} & 1.35 {[}1.16,1.56{]} & 0.86 & 1.46 {[}1.31,1.62{]} & 1.48 {[}1.34,1.63{]} & 0.89 \\
          & 75-79 & 1.26 {[}1.12,1.43{]} & 1.30 {[}1.13,1.48{]} & 0.86 & 1.22 {[}1.11,1.34{]} & 1.26 {[}1.14,1.38{]} & 0.80 \\
          & 80-84 & 1.13 {[}1.02,1.25{]} & 1.18 {[}1.06,1.31{]} & 0.77 & 1.15 {[}1.07,1.24{]} & 1.18 {[}1.10,1.26{]} & 0.75 \\
          & 85-89 & 1.11 {[}0.97,1.26{]} & 1.16 {[}1.02,1.31{]} & 0.79 & 1.07 {[}0.96,1.20{]} & 1.16 {[}1.06,1.28{]} & 0.54 \\
          & 90-99 & 1.30 {[}1.04,1.63{]} & 1.26 {[}1.04,1.52{]} & 0.86 & 1.11 {[}0.92,1.34{]} & 1.18 {[}1.04,1.33{]} & 0.71 \\
\bottomrule
\end{tabular}
\end{table}

\end{landscape}

\hypertarget{deviance-information-criterion}{%
\subsubsection{Deviance information criterion}\label{deviance-information-criterion}}

For each model in \textbf{Table \ref{tab:model-specification-summary}} we estimated
the deviance information criterion. The DIC was estimated using equation 7.9
from Gelman et al.\citep[see][ \S7]{Gelman2013} via the expected values of; the hazard
ratios, the model parameters, and the log-likelihood. These expectations were
taken over the posterior of the model parameters, \(\Theta | X\), and this
operator was approximated using the posterior samples, denoted \({HR}_j\) and
\(Z_j\), obtained by Markov Chain Monte Carlo. Thus the expected values are
denoted \(\hat{HR} = \sum_j {HR}_j / n_{\text{sample}}\), \(\hat{Z} = \sum_j Z_j / n_{\text{sample}}\), and \(\hat{\mathcal{L}} = \sum_j \mathcal{L}_{X | \Theta} \left( x | \Theta=\begin{bmatrix} {HR}_j & Z_j \end{bmatrix} \right) / n_{\text{sample}}\), respectively for the hazard
ratios, model parameters and log-likelihood. The DIC was then approximated
using the following equation:

\[
    DIC = -4 \hat{\mathcal{L}} +
        2 \mathcal{L}_{X | \Theta} \left(
            x | \Theta=\begin{bmatrix} \hat{HR} & \hat{Z} \end{bmatrix}
        \right).
\]

The estimate of the DIC for each model is shown in \textbf{Table
\ref{tab:model-deviance-information-criterion}} by sex.

\begin{table}[ht]
\centering
\caption{\label{tab:model-deviance-information-criterion}Estimated deviance information criterion (Gelman et al, 2013 \citep[see][pp.~172-173]{Gelman2013}) for each candidate model in the main analysis.}
\begin{tabular}{
  >{\raggedright\arraybackslash\bfseries}p{(\columnwidth - 8\tabcolsep) * \real{0.10}}
  >{\raggedright\arraybackslash}p{(\columnwidth - 8\tabcolsep) * \real{0.66}}
  >{\raggedleft\arraybackslash}p{(\columnwidth - 8\tabcolsep) * \real{0.12}}
  >{\raggedleft\arraybackslash}p{(\columnwidth - 8\tabcolsep) * \real{0.12}}}
\toprule
Model & \bfseries Description & \bfseries Women & \bfseries Men \\
\midrule
Null & Constant proportion that initiate, constant quit rate, and no switching to never smoker for those who quit. & 24,673.8 & 27,894.2 \\
A & As `Null' +2 d.f. to birth year spline-effect in proportion that initiated. & 24,134.3 & 24,516.2 \\
B & As `A' +2 d.f. to age spline-effect in quit rate. & 23,512.1 & 22,633.2 \\
C & As `B' +1 d.f. to calendar year spline-effect in quit rate. & 20,020.6 & 20,493.3 \\
D & As `C' and allowed those who quit before age 40 years to switch to never smoker. & 19,243.6 & 19,703.0 \\
E & As `D' +1 d.f. to calendar year spline-effect in quit rate. & 19,178.3 & 19,571.5 \\
F & As `E' +1 d.f. to birth year spline-effect in proportion that initiated. & 18,737.2 & 19,531.9 \\
G & As `F' +1 d.f. to birth year spline-effect in proportion that initiated. & 18,758.0 & 19,531.3 \\
\bottomrule
\end{tabular}
\end{table}

\hypertarget{model-discrepancy}{%
\subsubsection{Model discrepancy}\label{model-discrepancy}}

We used a Generalised Additive Model (GAM) to estimate the discrepancy, whose
expected value was assumed to be a Gaussian process\citep{Kammann2003} of the tensor
product of age and birth year spline effects. The covariance function of the
process was a squared exponential with smoothing parameters estimated by
Generalised Cross Validation implemented by the \texttt{mgcv} package in R.\citep{Wood2006, Wood2017} Three values were modelled in this fashion;

\begin{enumerate}
\def\labelenumi{\arabic{enumi}.}
\tightlist
\item
  The proportion of smokers in the population from 1962-2016, using all
  surveys.
\item
  The proportion of never-smokers amongst the non-smokers in the population
  from 1974-2016, excluding the AGP series in which non-smokers could not be
  differentiated further.
\item
  The proportion of those who quit before age 30 amongst ex-smokers in the
  population from 1980-2016 excluding the AGP and CCV series, and the
  NDSHS from 1985-95, which did not supply age-at-quit information.
\end{enumerate}

The mean and standard deviation of the predicted (expected) discrepancy, on a
log-odds scale, for the surveyed sample is shown in \textbf{Table
\ref{tab:discrepancy-in-proportions}}, where the expectation was taken over
the parameter posterior.

\begin{table}[ht]
\centering
\caption{\label{tab:discrepancy-in-proportions}Summary of the expected discrepancy in the log-odds of: being a smoker; being a never smoker if not a current smoker; and having quit before age 30 years if an ex-smoker, using Generalised Additive Models for the discrepancy fitted to the smoking survey data and the sample of the posterior expected proportions. Summary statistics shown are the estimated mean \(\bar x\) and standard deviation \(\bar s\) of the discrepancy, for each model (labels Null and A-G are described in \textbf{Table \ref{tab:model-specification-summary}}) and separately for men and women.}
\begin{tabular}{>{\bfseries}llrrrrrr}
\toprule
Sex & \bfseries Model & \(\bar x\) & \(\bar s\) & \(\bar x\) & \(\bar s\) & \(\bar x\) & \(\bar s\) \\
\midrule
Women & Null & -0.078 & 0.367 & -0.122 & 0.428 & -0.844 & 0.503 \\
      & A & -0.071 & 0.357 & -0.123 & 0.411 & -0.843 & 0.503 \\
      & B & -0.037 & 0.371 & -0.101 & 0.402 & -0.698 & 0.473 \\
      & C & -0.002 & 0.191 & -0.063 & 0.310 & -0.353 & 0.367 \\
      & D & -0.008 & 0.179 & 0.003 & 0.190 & -0.125 & 0.401 \\
      & E & -0.012 & 0.173 & -0.002 & 0.180 & -0.126 & 0.400 \\
      & F & -0.006 & 0.124 & -0.009 & 0.098 & -0.119 & 0.405 \\
      & G & -0.010 & 0.128 & -0.008 & 0.102 & -0.121 & 0.404 \\
Men & Null & -0.145 & 0.498 & -0.170 & 0.330 & -1.019 & 0.435 \\
    & A & -0.106 & 0.279 & -0.177 & 0.377 & -1.016 & 0.434 \\
    & B & -0.055 & 0.284 & -0.166 & 0.356 & -0.612 & 0.411 \\
    & C & -0.004 & 0.160 & -0.099 & 0.315 & -0.389 & 0.322 \\
    & D & -0.007 & 0.143 & -0.019 & 0.184 & -0.110 & 0.346 \\
    & E & -0.011 & 0.120 & -0.025 & 0.171 & -0.101 & 0.332 \\
    & F & -0.010 & 0.117 & -0.023 & 0.153 & -0.106 & 0.331 \\
    & G & -0.010 & 0.115 & -0.024 & 0.152 & -0.104 & 0.333 \\
\bottomrule
\end{tabular}
\end{table}

\clearpage

\hypertarget{additional-tables-and-figures}{%
\subsection{Additional tables and figures}\label{additional-tables-and-figures}}

\begin{table}[ht]
\centering
\caption{\label{tab:calibrated-change-in-ever-smoker}Sample obtained from the selected model of the growth in the proportion of Australians that initiated daily smoking by sex (men, women and persons) and ten-year increments of birth year from 1910 to 1990 and birth year 1996. Shown are the sample median and the interval given by the 5\textsuperscript{th} and 95\textsuperscript{th} percentiles of each sample. p-value: M $\equiv$ W; the mean proportion of more extreme values for men compared to each value for women.}
\begin{tabular}[]{
  >{\raggedright\arraybackslash}p{(\columnwidth - 10\tabcolsep) * \real{0.13}}
  >{\raggedleft\arraybackslash}p{(\columnwidth - 10\tabcolsep) * \real{0.24}}
  >{\raggedleft\arraybackslash}p{(\columnwidth - 10\tabcolsep) * \real{0.24}}
  >{\raggedleft\arraybackslash}p{(\columnwidth - 10\tabcolsep) * \real{0.24}}
  >{\raggedleft\arraybackslash}p{(\columnwidth - 10\tabcolsep) * \real{0.10}}}
\toprule
\bfseries Birth year & \bfseries Women & \bfseries Men & \bfseries Persons & \bfseries p-value: M $\equiv$ W \\
\midrule
1910 & -0.25\% {[}-0.46\%,-0.02\%{]} & -0.48\% {[}-0.51\%,-0.45\%{]} & -0.37\% {[}-0.47\%,-0.26\%{]} & 0.068 \\
1920 & -0.11\% {[}-0.30\%,0.10\%{]} & -0.64\% {[}-0.69\%,-0.58\%{]} & -0.38\% {[}-0.47\%,-0.28\%{]} & $\leq$0.001 \\
1930 & 0.32\% {[}0.19\%,0.46\%{]} & -0.75\% {[}-0.82\%,-0.69\%{]} & -0.23\% {[}-0.29\%,-0.16\%{]} & $\leq$0.001 \\
1940 & 0.97\% {[}0.90\%,1.04\%{]} & -0.76\% {[}-0.81\%,-0.71\%{]} & 0.08\% {[}0.05\%,0.12\%{]} & $\leq$0.001 \\
1950 & 1.00\% {[}0.91\%,1.09\%{]} & -0.80\% {[}-0.87\%,-0.74\%{]} & 0.08\% {[}0.03\%,0.13\%{]} & $\leq$0.001 \\
1960 & 0.17\% {[}0.10\%,0.23\%{]} & -1.11\% {[}-1.19\%,-1.04\%{]} & -0.48\% {[}-0.53\%,-0.44\%{]} & $\leq$0.001 \\
1970 & -1.53\% {[}-1.60\%,-1.46\%{]} & -1.88\% {[}-1.98\%,-1.80\%{]} & -1.71\% {[}-1.77\%,-1.66\%{]} & $\leq$0.001 \\
1980 & -3.82\% {[}-4.04\%,-3.59\%{]} & -2.92\% {[}-3.14\%,-2.70\%{]} & -3.36\% {[}-3.52\%,-3.23\%{]} & $\leq$0.001 \\
1990 & -6.12\% {[}-6.52\%,-5.71\%{]} & -3.90\% {[}-4.30\%,-3.54\%{]} & -4.98\% {[}-5.25\%,-4.73\%{]} & $\leq$0.001 \\
1996 & -7.03\% {[}-7.50\%,-6.56\%{]} & -4.33\% {[}-4.80\%,-3.92\%{]} & -5.65\% {[}-5.95\%,-5.36\%{]} & $\leq$0.001 \\
\bottomrule
\end{tabular}
\end{table}

\begin{table}[ht]
\centering
\caption{\label{tab:calibrated-quit-rate-growth}Sample obtained from the selected model of the growth in the quit rate of Australian smokers by sex, and ten-year increments of calendar year from 1930 to 2010 and in 2016. Shown are the sample median and the interval given by the 5\textsuperscript{th} and 95\textsuperscript{th} percentiles of each sample. p-value: M $\equiv$ W; the mean proportion of more extreme values for men compared to each value for women.}
\begin{tabular}{lrrr}
\toprule
Year & \bfseries Women & \bfseries Men & \bfseries p-value: M $\equiv$ W \\
\midrule
1930 & 6.36\% {[}5.60\%,7.17\%{]} & 5.32\% {[}4.71\%,5.83\%{]} & 0.092 \\
1940 & 6.22\% {[}5.49\%,7.00\%{]} & 5.18\% {[}4.60\%,5.67\%{]} & 0.082 \\
1950 & 5.82\% {[}5.19\%,6.52\%{]} & 4.79\% {[}4.29\%,5.22\%{]} & 0.053 \\
1960 & 5.17\% {[}4.68\%,5.71\%{]} & 4.14\% {[}3.77\%,4.49\%{]} & 0.019 \\
1970 & 4.26\% {[}3.95\%,4.64\%{]} & 3.24\% {[}2.99\%,3.48\%{]} & 0.000 \\
1980 & 3.24\% {[}3.06\%,3.43\%{]} & 2.22\% {[}2.10\%,2.34\%{]} & $\leq$0.001 \\
1990 & 2.44\% {[}2.30\%,2.62\%{]} & 1.43\% {[}1.25\%,1.59\%{]} & $\leq$0.001 \\
2000 & 1.91\% {[}1.65\%,2.18\%{]} & 0.90\% {[}0.64\%,1.13\%{]} & $\leq$0.001 \\
2010 & 1.61\% {[}1.29\%,1.95\%{]} & 0.61\% {[}0.32\%,0.88\%{]} & $\leq$0.001 \\
2016 & 1.56\% {[}1.23\%,1.91\%{]} & 0.55\% {[}0.26\%,0.84\%{]} & $\leq$0.001 \\
\bottomrule
\end{tabular}
\end{table}

\begin{table}[ht]
\centering
\caption{\label{tab:calibrated-quit-before-twenty}Sample obtained from the selected model of the proportion who quit daily smoking prior to age 20 years amongst Australians who had initiated smoking and were born between 1910 and 1996, shown by sex. Shown are the sample median and the interval given by the 5\textsuperscript{th} and 95\textsuperscript{th} percentiles of each sample.}
\begin{tabular}{rr}
\toprule
\bfseries Women & \bfseries Men \\
\midrule
27.0\% {[}26.7\%,27.3\%{]} & 27.0\% {[}26.8\%,27.3\%{]} \\
\bottomrule
\end{tabular}
\end{table}

\clearpage

\hypertarget{sensitivity-analysis}{%
\subsubsection{Sensitivity analysis}\label{sensitivity-analysis}}

\begin{table}[ht]
\centering
\caption{\label{tab:sensitivity-analysis-DIC}Estimated DIC of each alternative model calibrated in sensitivity analyses, excluding those that are not comparable to the main analysis (the alternative of no survey weights, and the alternative initial age of 25 years).}
\begin{tabular}{
  >{\raggedright\arraybackslash}p{(\columnwidth - 6\tabcolsep) * \real{0.80}}
  >{\raggedleft\arraybackslash}p{(\columnwidth - 6\tabcolsep) * \real{0.10}}
  >{\raggedleft\arraybackslash}p{(\columnwidth - 6\tabcolsep) * \real{0.10}}}
\toprule
\bfseries Description & \bfseries Women & \bfseries Men \\
\midrule
Selected model & 18737.18 & 19531.92 \\
Used hazard ratios observed in Cancer Prevention Study I rather than 45 and Up Study. & 18863.21 & 19593.44 \\
Allowed those who quit after age 40 years to switch to never smoker. & 18723.58 & 19533.04 \\
No ex-smokers were able to switch to never smoker. & 19373.40 & 20167.48 \\
Used cohort instead of calendar year spline-effect in quit rate. & 18801.83 & 19638.50 \\
\bottomrule
\end{tabular}
\end{table}

\begin{table}[ht]
\centering
\caption{\label{tab:sensitivity-analysis-proportion-initiate}Sample obtained from the selected model and the sensitivity analysis models of the proportion in an Australian birth cohort that initiated daily smoking for birth year 1996 by sex. Shown are the sample median and the interval given by the 5\textsuperscript{th} and 95\textsuperscript{th} percentiles of each sample.}
\begin{tabular}{
  >{\raggedright\arraybackslash}p{(\columnwidth - 6\tabcolsep) * \real{0.68}}
  >{\raggedleft\arraybackslash}p{(\columnwidth - 6\tabcolsep) * \real{0.16}}
  >{\raggedleft\arraybackslash}p{(\columnwidth - 6\tabcolsep) * \real{0.16}}}
\toprule
\bfseries Description & \bfseries Women & \bfseries Men \\
\midrule
Selected model & 16.3\% {[}15.2\%,17.5\%{]} & 22.7\% {[}21.1\%,24.2\%{]} \\
Used hazard ratios observed in Cancer Prevention Study I rather than 45 and Up Study. & 16.2\% {[}15.2\%,17.1\%{]} & 22.5\% {[}21.1\%,24.1\%{]} \\
Used cohort instead of calendar year spline-effect in quit rate. & 16.4\% {[}15.5\%,17.3\%{]} & 22.6\% {[}21.2\%,24.1\%{]} \\
Allowed those who quit after age 40 years to switch to never smoker. & 16.5\% {[}15.5\%,17.5\%{]} & 22.5\% {[}21.1\%,24.1\%{]} \\
No ex-smokers were able to switch to never smoker. & 17.5\% {[}16.4\%,18.6\%{]} & 23.2\% {[}21.9\%,24.6\%{]} \\
Ignored weights for smoking survey responses in likelihood. & 18.7\% {[}17.8\%,19.8\%{]} & 24.0\% {[}22.8\%,25.3\%{]} \\
\bottomrule
\end{tabular}
\end{table}

\begin{table}[ht]
\centering
\caption{\label{tab:first-age-analysis-proportion-initiate}Sample obtained from the selected model and also a sensitivity analysis, with the age at completed initiation was raised to 25 years, of the proportion in an Australian birth cohort that initiated daily smoking for birth year 1991 by sex. Shown are the sample median and the interval given by the 5\textsuperscript{th} and 95\textsuperscript{th} percentiles of each sample.}
\begin{tabular}{
  >{\raggedright\arraybackslash}p{(\columnwidth - 6\tabcolsep) * \real{0.56}}
  >{\raggedleft\arraybackslash}p{(\columnwidth - 6\tabcolsep) * \real{0.22}}
  >{\raggedleft\arraybackslash}p{(\columnwidth - 6\tabcolsep) * \real{0.22}}}
\toprule
\bfseries Description & \bfseries Women & \bfseries Men \\
\midrule
Selected model & 22.8\% {[}21.8\%,24.0\%{]} & 27.9\% {[}26.6\%,29.2\%{]} \\
Raised the age when the model starts to 25 years. & 22.9\% {[}21.6\%,24.1\%{]} & 32.0\% {[}30.1\%,34.3\%{]} \\
\bottomrule
\end{tabular}
\end{table}

\afterpage{

\begin{landscape}
\thispagestyle{lscape}
\pagestyle{lscape}

\begin{table}[t]
\centering
\caption{\label{tab:sensitivity-analysis-quit-rate}Sample obtained from the selected model and the sensitivity analyses of the rate that Australian smokers quit daily smoking (per 100 person-years) in the calendar year 2016 by age at 30, 50 and 70 years, and by sex. Shown are the sample median and the interval given by the 5\textsuperscript{th} and 95\textsuperscript{th} percentiles of each sample.}
\begin{tabular}{
  >{\raggedright\arraybackslash}p{(\columnwidth - 14\tabcolsep) * \real{0.46}}
  >{\raggedleft\arraybackslash}p{(\columnwidth - 14\tabcolsep) * \real{0.09}}
  >{\raggedleft\arraybackslash}p{(\columnwidth - 14\tabcolsep) * \real{0.09}}
  >{\raggedleft\arraybackslash}p{(\columnwidth - 14\tabcolsep) * \real{0.09}}
  >{\raggedleft\arraybackslash}p{(\columnwidth - 14\tabcolsep) * \real{0.09}}
  >{\raggedleft\arraybackslash}p{(\columnwidth - 14\tabcolsep) * \real{0.09}}
  >{\raggedleft\arraybackslash}p{(\columnwidth - 14\tabcolsep) * \real{0.09}}}
\toprule
  &
  \multicolumn{2}{c}{\bfseries Age 30 years} &
  \multicolumn{2}{c}{\bfseries Age 50 years} &
  \multicolumn{2}{c}{\bfseries Age 70 years} \\
                  \cmidrule(lr){2-3} \cmidrule(lr){4-5} \cmidrule(l){6-7}
\bfseries Description &
    \bfseries Women & \bfseries Men &
    \bfseries Women & \bfseries Men &
    \bfseries Women & \bfseries Men \\
\cmidrule(r){1-1} \cmidrule(lr){2-3} \cmidrule(lr){4-5} \cmidrule(l){6-7}
Selected model & 5.52 {[}5.19,5.88{]} & 3.67 {[}3.44,3.92{]} & 5.00 {[}4.77,5.31{]} & 4.72 {[}4.43,4.99{]} & 6.05 {[}5.36,6.72{]} & 4.70 {[}4.13,5.29{]} \\
Used hazard ratios observed in Cancer Prevention Study I rather than 45 and Up Study. & 5.52 {[}5.14,5.83{]} & 3.62 {[}3.41,3.81{]} & 5.15 {[}4.85,5.45{]} & 4.81 {[}4.56,5.08{]} & 7.90 {[}7.30,8.51{]} & 6.53 {[}6.11,7.10{]} \\
Raised the age when the model starts to 25 years. & 5.32 {[}5.07,5.57{]} & 3.80 {[}3.59,4.02{]} & 4.62 {[}4.39,4.81{]} & 4.53 {[}4.33,4.69{]} & 5.60 {[}5.08,6.22{]} & 4.82 {[}4.38,5.31{]} \\
Used cohort instead of calendar year spline-effect in quit rate. & 6.65 {[}6.33,6.97{]} & 4.68 {[}4.38,4.97{]} & 5.97 {[}5.71,6.20{]} & 5.96 {[}5.74,6.18{]} & 7.81 {[}7.18,8.54{]} & 6.83 {[}6.21,7.46{]} \\
Allowed those who quit after age 40 years to switch to never smoker. & 5.45 {[}5.15,5.79{]} & 3.61 {[}3.42,3.85{]} & 5.23 {[}4.96,5.54{]} & 4.69 {[}4.47,4.93{]} & 7.02 {[}6.35,7.72{]} & 4.81 {[}4.22,5.40{]} \\
No ex-smokers were able to switch to never smoker. & 5.21 {[}4.94,5.63{]} & 3.34 {[}3.14,3.57{]} & 5.42 {[}5.16,5.87{]} & 4.90 {[}4.69,5.19{]} & 6.84 {[}6.15,7.44{]} & 4.59 {[}4.09,5.12{]} \\
Ignored weights for smoking survey responses in likelihood. & 5.51 {[}5.20,5.82{]} & 3.42 {[}3.27,3.65{]} & 5.11 {[}4.89,5.37{]} & 4.50 {[}4.30,4.71{]} & 6.57 {[}5.97,7.17{]} & 4.80 {[}4.37,5.35{]} \\
\bottomrule
\end{tabular}
\end{table}

\begin{table}[t]
\centering
\caption{\label{tab:sensitivity-analysis-reversion-rate}Samples obtained from the selected model and the sensitivity analyses of the rate that Australian ex-smokers switch to reporting as never (per 100 person-years) by age-at-quit group and sex. Shown are the sample median and the interval given by the 5\textsuperscript{th} and 95\textsuperscript{th} percentiles of each sample.}
\begin{tabular}[]{@{}
  >{\raggedright\arraybackslash}p{(\columnwidth - 12\tabcolsep) * \real{0.46}}
  >{\raggedleft\arraybackslash}p{(\columnwidth - 12\tabcolsep) * \real{0.09}}
  >{\raggedleft\arraybackslash}p{(\columnwidth - 12\tabcolsep) * \real{0.09}}
  >{\raggedleft\arraybackslash}p{(\columnwidth - 12\tabcolsep) * \real{0.09}}
  >{\raggedleft\arraybackslash}p{(\columnwidth - 12\tabcolsep) * \real{0.09}}
  >{\raggedleft\arraybackslash}p{(\columnwidth - 12\tabcolsep) * \real{0.09}}
  >{\raggedleft\arraybackslash}p{(\columnwidth - 12\tabcolsep) * \real{0.09}}@{}}
\toprule
  &
  \multicolumn{2}{c}{\bfseries Age 30 years} &
  \multicolumn{2}{c}{\bfseries Age 50 years} &
  \multicolumn{2}{c}{\bfseries Age 70 years} \\
                  \cmidrule(lr){2-3} \cmidrule(rl){4-5} \cmidrule(l){6-7}
\bfseries Description &
    \bfseries Women & \bfseries Men &
    \bfseries Women & \bfseries Men &
    \bfseries Women & \bfseries Men \\
\cmidrule(r){1-1} \cmidrule(lr){2-3} \cmidrule(lr){4-5} \cmidrule(l){6-7}
Selected model & 2.31 {[}2.14,2.46{]} & 2.07 {[}1.94,2.23{]} & 0.86 {[}0.66,1.03{]} & 0.31 {[}0.15,0.45{]} & - & - \\
Used hazard ratios observed in Cancer Prevention Study I rather than 45 and Up Study. & 2.34 {[}2.18,2.50{]} & 2.06 {[}1.92,2.20{]} & 0.80 {[}0.60,1.02{]} & 0.19 {[}0.07,0.34{]} & - & - \\
Raised the age when the model starts to 25 years. & 2.20 {[}2.02,2.36{]} & 1.95 {[}1.80,2.10{]} & 1.06 {[}0.83,1.25{]} & 0.49 {[}0.34,0.64{]} & - & - \\
Used cohort instead of calendar year spline-effect in quit rate. & 2.45 {[}2.33,2.60{]} & 2.24 {[}2.12,2.39{]} & 0.92 {[}0.73,1.11{]} & 0.30 {[}0.18,0.47{]} & - & - \\
Allowed those who quit after age 40 years to switch to never smoker. & 2.27 {[}2.12,2.42{]} & 2.08 {[}1.94,2.18{]} & 0.87 {[}0.70,1.07{]} & 0.26 {[}0.13,0.43{]} & 0.71 {[}0.43,0.99{]} & 0.04 {[}0.00,0.19{]} \\
Ignored weights for smoking survey responses in likelihood. & 2.42 {[}2.30,2.54{]} & 2.06 {[}1.95,2.19{]} & 0.67 {[}0.48,0.83{]} & 0.17 {[}0.06,0.29{]} & - & - \\
\bottomrule
\end{tabular}
\end{table}

\begin{table}[t]
\centering
\caption{\label{tab:hazard-ratio-in-sensitivity-analysis}Age-standardised sex-specific hazard ratio of death from all causes for smokers and ex-smokers compared to never smokers in each sensitivity analyses with prior hazard that were estimated from the 45 and Up Study. Hazard ratios were standardised to the sex-specific distribution of age in the Australian population from calendar year 2000 sourced from Human Mortality Database \citep{HMD2019}. Shown are; the sample median and the interval given by the 5\textsuperscript{th} and 95\textsuperscript{th} percentiles of each sample in the sensitivity analyses and the selected model of the main analysis; the same percentiles for the prior distribution; and the estimated overlap statistics compared to the prior distribution.}
\begin{tabular}{
  >{\raggedright\arraybackslash\bfseries}p{(\columnwidth - 12\tabcolsep) * \real{0.07}}
  >{\raggedright\arraybackslash}p{(\columnwidth - 12\tabcolsep) * \real{0.49}}
  >{\raggedleft\arraybackslash}p{(\columnwidth - 12\tabcolsep) * \real{0.14}}
  >{\raggedleft\arraybackslash}p{(\columnwidth - 12\tabcolsep) * \real{0.08}}
  >{\raggedleft\arraybackslash}p{(\columnwidth - 12\tabcolsep) * \real{0.14}}
  >{\raggedleft\arraybackslash}p{(\columnwidth - 12\tabcolsep) * \real{0.08}}}
\toprule
& &
    \multicolumn{2}{c}{\bfseries Smoker} &
    \multicolumn{2}{c}{\bfseries Ex-smoker} \\
                  \cmidrule(lr){3-4} \cmidrule(l){5-6}
Sex & \bfseries Description &
    \bfseries Hazard ratio & \bfseries Overlap &
    \bfseries Hazard ratio & \bfseries Overlap \\
\cmidrule(r){1-2} \cmidrule(lr){3-4} \cmidrule(l){5-6}
Women & Prior (45 and Up Study) & 3.23 {[}2.96,3.54{]} & - & 1.36 {[}1.26,1.48{]} & - \\
      & Selected model & 3.45 {[}3.17,3.79{]} & 0.54 & 1.37 {[}1.27,1.50{]} & 0.89 \\
      & Raised the age when the model starts to 25 years. & 3.49 {[}3.23,3.78{]} & 0.48 & 1.37 {[}1.26,1.48{]} & 0.95 \\
      & Used cohort instead of calendar year spline-effect in quit rate. & 3.40 {[}3.13,3.70{]} & 0.65 & 1.38 {[}1.27,1.50{]} & 0.89 \\
      & Allowed those who quit after age 40 years to switch to never smoker. & 3.31 {[}3.08,3.57{]} & 0.81 & 1.34 {[}1.26,1.44{]} & 0.86 \\
      & No ex-smokers were able to switch to never smoker. & 3.56 {[}3.26,3.83{]} & 0.41 & 1.50 {[}1.38,1.66{]} & 0.37 \\
      & Ignored weights for smoking survey responses in likelihood. & 3.34 {[}3.09,3.65{]} & 0.73 & 1.35 {[}1.26,1.46{]} & 0.96 \\
Men & Prior (45 and Up Study) & 3.60 {[}3.31,3.93{]} & - & 1.31 {[}1.21,1.44{]} & - \\
    & Selected model & 3.51 {[}3.21,3.83{]} & 0.79 & 1.31 {[}1.21,1.42{]} & 0.94 \\
    & Raised the age when the model starts to 25 years. & 3.53 {[}3.26,3.81{]} & 0.83 & 1.30 {[}1.20,1.43{]} & 0.93 \\
    & Used cohort instead of calendar year spline-effect in quit rate. & 3.46 {[}3.22,3.81{]} & 0.72 & 1.31 {[}1.21,1.42{]} & 0.95 \\
    & Allowed those who quit after age 40 years to switch to never smoker. & 3.46 {[}3.18,3.76{]} & 0.73 & 1.29 {[}1.21,1.40{]} & 0.86 \\
    & No ex-smokers were able to switch to never smoker. & 3.62 {[}3.42,3.92{]} & 0.91 & 1.43 {[}1.33,1.56{]} & 0.38 \\
    & Ignored weights for smoking survey responses in likelihood. & 3.31 {[}3.05,3.56{]} & 0.41 & 1.29 {[}1.19,1.41{]} & 0.86 \\
\bottomrule
\end{tabular}
\end{table}

\end{landscape}

} 

\clearpage

\hypertarget{informal-validation-of-quit-rate}{%
\subsubsection{Informal validation of quit rate}\label{informal-validation-of-quit-rate}}

\begin{figure}[ht]
\centering
\includegraphics{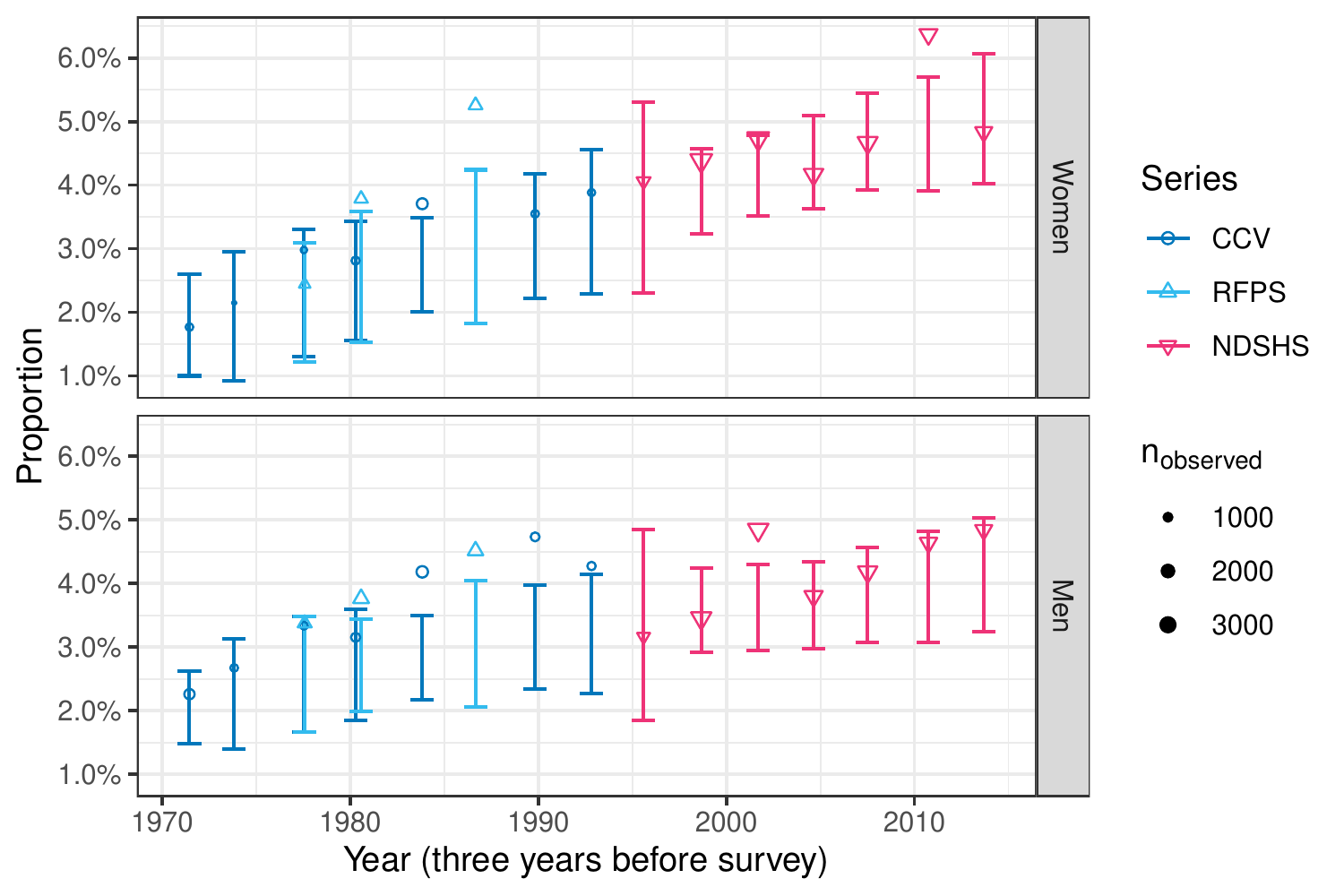}
\caption{\label{fig:cross-section-quit-rate}Proportion of Australian smokers who had quit over a one-year period. Shown; the estimate obtained from cross-sectional surveys (markers) using the proportion who were abstinent for at least two years at the time of the survey amongst those who still smoked or had stopped within the five years prior to being surveyed; and the interval given by the 5\textsuperscript{th} and 95\textsuperscript{th} percentiles of the sampled predictions of the proportion obtained from the selected model for each survey (error bars). Size of marker indicates number of respondents who smoked between two- and five- years prior.Abbreviations; CCV: Cancer Council Victoria adult smoking survey; RFPS: Risk Factor Prevalence Study; NDSHS: National Drug Strategy Household Survey.}
\end{figure}

We defined a quit event as the event that a smoker quits and never again smokes
(on a daily basis) and in the cross-sectional survey data these events were not
distinguishable from `attempts' at quitting with later relapse; therefore we
cannot validate the estimate of the model's quit rate against the proportion of
those who smoked within the last year who had maintained abstinence up until the
survey. However, in the ITC Four Country Survey, a longitudinal study, it was
observed that 95\% of those with at least two years abstinence maintained
abstinence over the next year \citep{Herd2009}. We estimated the proportion of
smokers that had quit over a one year period (three years prior to the survey)
using the proportion in the survey who had maintained abstinence for at least
two years amongst those who either still smoked or had quit within the five
years prior to being surveyed, and compared this to the prediction from the
model. The survey-estimated proportion is shown in \textbf{Figure
\ref{fig:cross-section-quit-rate}} along with the interval given by the
5\textsuperscript{th} and 95\textsuperscript{th} percentiles of the sampled predictions
obtained from the selected model. Most of the survey-estimated values lie within
the intervals, with all outliers being in excess of the upper limit of the
interval, which is expected given that the survey estimate would be an
over-estimate due to the occurrence of relapse.

\end{document}